\documentclass[prd,showpacs,preprintnumbers,amsmath,amssymb,floatfix]{revtex4}

\usepackage{graphicx}
\usepackage{dcolumn}
\usepackage{bm}
\usepackage{amssymb}
\usepackage{epsfig}
\usepackage{color}

\newcommand{\ba}{\begin{eqnarray}}
\newcommand{\ea}{\end{eqnarray}}
\newcommand{\be}{\begin{equation}}
\newcommand{\ee}{\end{equation}}
\newcommand{\bdisplay}{\begin{displaymath}}
\newcommand{\edisplay}{\end{displaymath}}

\newcommand{\eq}[1]{Eq.\,(\ref{#1})}
\newcommand{\fig}[1]{Fig.\,\ref{#1}}
\newcommand{\Sec}[1]{Sec.\,\ref{#1}}

\begin{document}

\title{Eikonal fit to $pp$ and $\bar{p}p$ scattering and the edge in the scattering amplitude  }

\author{Martin~M.~Block}
\email{mblock@northwestern.edu}
\affiliation{Department of Physics and Astronomy, Northwestern University,
Evanston, IL 60208}
\author{Loyal Durand}
\email{ldurand@hep.wisc.edu}
\altaffiliation{Mailing address: 415 Pearl Ct., Aspen, CO 81611}
\affiliation{Department of Physics, University of Wisconsin, Madison, WI 53706}
\author{Phuoc Ha}
\email{pdha@towson.edu}
\affiliation{Department of Physics, Astronomy and Geosciences, Towson University, Towson, MD 21252}
\author{Francis Halzen}
\email{francis.halzen@icecube.wisc.edu}
\affiliation{Wisconsin IceCube Particle Astrophysics Center and Department
of Physics, University of Wisconsin-Madison,  Madison,  Wisconsin 53706}

\begin{abstract}
We make a detailed eikonal fit to current data on the total and elastic scattering cross sections, the ratios $\rho$ of the real to the imaginary parts of the forward elastic scattering amplitudes, and the logarithmic slopes $B$ of the differential cross sections $d\sigma/dt$ at $t=0$, for proton-proton and antiproton-proton scattering at center-of-mass energies $W$ from 5 GeV to 57 TeV. The fit allows us to investigate the structure of the eikonal amplitudes in detail, including the impact-parameter structure of the energy-independent edge in the scattering amplitude shown to exist by Block {\em et al.} \cite{edge}. We show that the edge region has an essentially fixed shape with a peak at approximately the ``black disk'' radius $R_{\rm tot}=\sqrt{\sigma_{\rm tot}/2\pi}$ of the scattering amplitude, a constant width $t_{\rm edge}\approx 1$ fm, and migrates to larger impact parameters with increasing energy proportionally to $R_{\rm tot}$. We comment on possible physical mechanisms which could lead to the edge. We show that the eikonal results for the cross sections  and $\rho$ values are  described to high accuracy by analytic expressions of the forms used in earlier analyses by Block and Halzen, and extend the result to  the elastic-scattering slope parameter $B$. These expressions provide simple extrapolations of the results to much higher energies where the cross sections approach the black disk limit with $\sigma_{\rm elas},\ \sigma_{\rm inel}\rightarrow \sigma_{\rm tot}/2$ and $B\rightarrow\sigma_{\rm tot}/8\pi$. Finally, we calculate the survival probabilities for large rapidity gaps in the scattering.
\end{abstract}


\maketitle


\section{Introduction \label{introduction} }

In a recent paper \cite{edge} Block {\em et al.} established that the proton-proton scattering amplitude has an edge in impact parameter space with a width that remains essentially constant over many orders of magnitude in the center-of-mass energy $W$. This result was derived using  the general forms of the scattering amplitudes in impact parameter space for strongly absorptive scattering, and the very accurate Block-Halzen fit to the proton-proton ($pp$) and antiproton ($\bar{p}p$) total and inelastic cross sections and ratios $\rho$ of the real to imaginary parts of the forward elastic scattering amplitudes for $6\leq W\leq 1800$ GeV \cite{blockhalzenfit}. That fit incorporated the asymptotic $\ln^2s$ limit of the growth of the cross sections at large $s=W^2$, the constraints on the phase of the scattering amplitude imposed by analyticity and crossing symmetry, and constraints on its magnitude and slope at 4 GeV implied by consistency with low-energy data. It successfully predicted the cross sections subsequently measured in the multi-TeV range at the Large Hadron Collider (LHC) and in cosmic ray experiments \cite{blockhalzen,blockhalzen2}.

The results in \cite{edge} did not depend on a particular fit to the scattering amplitudes, but only on the general forms of the amplitudes in impact parameter space. In the present paper, we use a detailed eikonal description of the scattering to fit the $pp$ and $\bar{p}p$ data on $\sigma_{\rm tot},\ \sigma_{\rm elas},\  \rho,$ and on $B$, the logarithmic derivative of $d\sigma/dt$ at $t=0$.  The fit includes a number of new measurements at higher energies. While the results we obtain for these quantities are essentially equivalent to those obtained earlier, the detailed fit allows us the investigate the structure of the eikonal amplitudes including, in particular, the structure of the edge in impact parameter space.

In the following sections, we first establish our conventions and give expressions for the cross sections, $\rho$, and $B$ in terms of the eikonal function  (\Sec{sec:conventions}).   We use a general parametrization of the eikonal function which, importantly, incorporates the power-law growth $\propto s^\epsilon$ found in QCD-based minijet \cite{fagundes} and Reggeon \cite{KhozeMartinRyskin,GotsmanLHC} models, and the exponential cutoff in impact parameter suggested by the proton form factor. This results in the asymptotic approach of the scattering amplitudes to the black-disk limit in which $\sigma_{\rm tot}\propto \ln^2s$, $\sigma_{\rm elas},\,\sigma_{\rm inel}\rightarrow \sigma_{\rm tot}/2$, $B\rightarrow \sigma_{\rm tot}/8\pi$, and $\rho\rightarrow 0$.

We present the results of our fit to the  $pp$ and $\bar{p}p$ data in \Sec{sec:fit}. We then use the results to investigate the structure of the eikonal amplitudes (\Sec{subsec:structure}), including the relative importance of different contributions to the eikonal function and the (slow) approach of the scattering to asymptotic behavior dominated by gluon-related processes.

We show in \Sec{subsec:BHconnection} that the description of the total and elastic cross sections and $\rho$ values obtained in the eikonal model can be fitted to high accuracy by analytic expressions of the form used in the Block-Halzen fits to the data \cite{blockhalzenfit,blockhalzen,blockhalzen2}, justifying their assumptions. The resulting expressions for these quantities can be extrapolated reliably to higher energies. We also extend this analysis to $B$, and give results that may be useful in other contexts such as the analysis of cosmic ray cross sections. We prefer these extrapolations to those using the eikonal model for reasons we discuss.

We then investigate the structure of the edge region in the scattering amplitude (\Sec{subsec:edge}). We find that the edge maintains a nearly constant shape in impact parameter space, with a width which remains essentially constant at $\sim1$ fm up to the highest energies studied to date, and presumably to much higher asymptotic energies. We comment on some possible explanations of the edge and its form at large impact parameters in \Sec{subsec:edge_origin}; this is a problem that needs further study.

We note  in \Sec{subsec:edge_origin} that the component of the cross section associated with the edge gives the Pumplin bound \cite{pumplin} on single-particle diffraction dissociation in $pp$ or $\bar{p}p$ collisions, and suggest that experiments to test the bound would be useful.
Finally, in \Sec{subsec:rapidity_gaps}, we discuss and calculate the survival probabilities for large rapidity gaps in the scattering.
The details of our eikonal model are discussed in the Appendix.


\section{Conventions \label{sec:conventions}}

In the following, we will be concerned with proton-proton ($pp$) and proton-antiproton ($p\bar{p}$) scattering at high energies. We will neglect the (presumably small) effects of the nucleon spins, and describe the scattering amplitude and cross sections in an impact parameter or eikonal representation; this is valid at small angles when many partial waves contribute to the scattering and the (unitary) partial wave series can be converted to an integral over the impact parameter. We will write the resulting spin-independent eikonal scattering amplitude and differential elastic scattering amplitude  as
\ba
\label{f}
f(s,t) &=& i\int_0^\infty db\,b\left(1-e^{i\chi(b,s)}\right) J_0(b\sqrt{-t}), \\
\label{dsigma/dt}
\frac{d\sigma}{dt}(s,t) &=& \pi\left|f(s,t)\right|^2.
\ea
Here $s=W^2=4(p^2+m^2)$ is the square of the total energy  in the center of mass (c.m.) system, $p$ is the c.m. momentum of either incident particle, $b=j/p$ where $j$ is the partial-wave angular momentum, and $t=-2p^2(1-\cos\theta)$ is the invariant 4-momentum transfer for elastic scattering at the angle $\theta$.
We will define the eikonal function $\chi(b,s)$  as $\chi=\chi_R+i\chi_I$; note that some other papers use different conventions, {\em e.g.,} \cite{edge,blockrev}.

With these conventions, the elastic, total, and inelastic cross sections are
\ba
\label{sigma_el}
\sigma_{\rm elas} &=& 2\pi\int_0^\infty db\, b \left |1-e^{i\chi}\right |^2 = 2\pi\int_0^\infty db\, b \left(1-2\cos{\chi_R}\, e^{-\chi_I}+e^{-2\chi_I}\right), \\
\label{sigma_tot}
\sigma_{\rm tot}(s) &=&4\pi {\rm Im} f(s,0) = 4\pi \int_0^\infty db\, b \left (1-\cos{\chi_R}\,e^{-\chi_I}\right), \\
\label{sigma_inel}
\sigma_{\rm inel}(s) &=& \sigma_{\rm tot}-\sigma_{\rm elas} = 2\pi\int_0^\infty db\, b \left (1-e^{-2\chi_I}\right).
\ea

The ratio $\rho$ of the real to the imaginary part of the forward scattering amplitude and the logarithmic derivative $B$ of the differential elastic scattering cross section at $t=0$ are also frequently measured and will be used in our analysis. Here
\ba
\label{rho}
\rho &=& {\rm Re}\,f(s,0)/{\rm Im}\,f(s,0) = -\int_0^\infty db\,b e^{-\chi_I}\sin{\chi_R}\Big/\int_0^\infty db\,b\left(1-\cos{\chi_R}\,e^{-\chi_I}\right), \\
\label{Bdef}
B &=& \frac{d}{dt}\left[\ln\frac{d\sigma}{dt}(s,t)\right]_{t=0} \\
&=& \frac{1}{2}\left[\int_0^\infty db\,b^3\sin{\chi_R}e^{-\chi_I}\,\int_0^\infty db\,b\sin{\chi_R}e^{-\chi_I}  + \int_0^\infty db\,b^3\left(1-\cos{\chi_R}e^{-\chi_I}\right)\,\int_0^\infty db\,b\left(1-\cos{\chi_R}e^{-\chi_I}\right) \right]  \nonumber \\
\label{B}
&& \bigg/ \left[\left(\int_0^\infty db\,b\sin{\chi_R}e^{-\chi_I}\right)^2 + \left(\int_0^\infty db\,b\left(1-\cos{\chi_R}e^{-\chi_I}\right)\right) ^2 \right].
\ea
An accurate approximation for $B$ when the real part of the scattering amplitude is small is to set $\chi_R=0$. Then
\be
\label{Bapprox}
B \approx \frac{1}{2}\int_0^\infty db\,b^3\left(1-e^{-\chi_I}\right)\bigg/\int_0^\infty db\,b\left(1-e^{-\chi_I}\right).
\ee
We have used the exact expression in \eq{B} in fitting the experimental data, but note that the approximate expression would have been adequate.


\section{Fit to high energy proton-proton and antiproton-proton scattering data \label{sec:fit}}

The model  we have used in fitting the high evergy $pp$ and $p\bar{p}$ cross sections is a modification of the ``Aspen model'' of  Block {\em et al.\/} \cite{blockaspen,blockrev} which was motivated by the structure of the eikonal function found in QCD minijet models for the scattering. We will follow the notation used in those references even though the precise identification of the terms made there cannot really be maintained in a more general setting. We write the eikonal functions in terms of crossing-even and crossing-odd components, with
\ba
\label{chipbarp}
 \chi_{p\bar{p}}(b,W) &=& \left[\chi_E(b,W) +\chi_O(b,W)\right]/2,  \\
\label{chipp}
\chi_{pp}(b,W)  &=&  \left[ \chi_E(b,W) -\chi_O(b,W)\right]/2.
\ea
The even and odd functions are defined as
\ba
\label{chiEmodel}
\chi_E(b,W) &=& i\left[\sigma_{qq}(We^{-i\pi/4}) A(b,\mu_{qq}) + \sigma_{qg}(We^{-i\pi/4}) A(b,\mu_{qg}) + \sigma_{gg}(We^{-i\pi/4}) A(b,\mu_{gg})\right], \\
\label{chiOmodel}
\chi_O(b,W) &=& -C_5\Sigma_{gg}\left(\frac{m_0}{W}e^{i\pi/4}\right)^{2-2\alpha_1} A(b,\mu_{odd}),
\ea
where the  phases of the functions in Eqs.\ (\ref{chiEmodel}) and (\ref{chiOmodel}) are determined by the constraints imposed by analyticity and crossing symmetry \cite{blockcahn,blockrev}.

In these expressions, the factors $A(b,\mu)$ are overlap functions for the colliding hadrons and the ``cross sections'' $\sigma_{ij}$ are intended to describe the interactions between the the corresponding components $i$ and $j$ of the two particles chosen from the matter ($q$)  or gluon ($g$) fields. The details of the model are given in Appendix \ref{model}.

Our parametrization of $\chi$ is general and very flexible, including a leading power-law dependence $s^\epsilon$, additional  logarithmic and constant terms, and falling Regge-like terms in $s$. Our objective is to get a good fit to all the $pp$ and $\bar{p}p$ data up to the highest energies where measurement exist, and to then use the results to study the eikonal structure of the scattering amplitudes with immediate emphasis on the edge region \cite{edge}. In contrast to our relatively free parametrization of $\chi$, other recent parametrizations such as those in \cite{fagundes,KhozeMartinRyskin,GotsmanLHC} are based on specific dynamical models, and those papers emphasize the testing of those models through fits to the data.

\begin{figure}[htbp]
\includegraphics{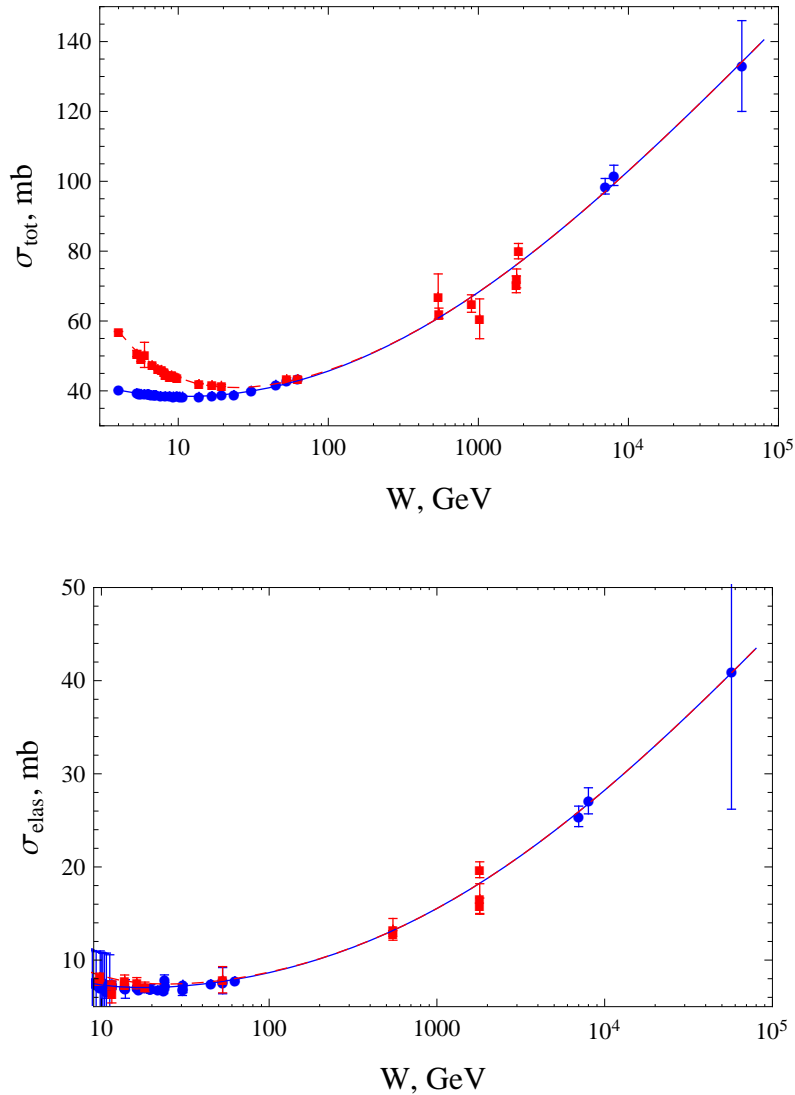}
\caption{Top panel: fits to $\sigma_{\rm tot,pp}$ (blue dots and solid line)  and $\sigma_{\rm tot,\bar{p}p}$ (red squares and dashed line).  Only data above 5 GeV were used in the final fit, with the cross sections constrained to fit compilations of low-energy data at 4 GeV \cite{blockrev}. . Bottom panel: fits to $\sigma_{\rm elas,pp}$ (blue dots and solid line)  and $\sigma_{\rm elas,\bar{p}p}$ (red squares and dashed line). The fit used only data above 10 GeV. }
\label{fig:crosssectionfits}
\end{figure}
%

We have used our parametrization and the expressions above to fit the combined data on $pp$ and $\bar{p}p$ total cross sections for $W\geq 5.3$ GeV and the elastic scattering cross sections, $\rho$, and $B$ for energies $W\geq 10$ GeV. The fit was further constrained as described in \cite{blockrev} by fixing the values of $\sigma_{\rm tot,pp}$ and $\sigma_{\rm tot,\bar{p}p}$ at $W=4$ GeV to match the results obtained from the extensive low-energy data. This is the same energy range with the same constraints as used in the Block-Halzen fits based on analytic amplitudes with a $\ln^2s$ high energy behavior \cite{blockhalzenfit,blockrev}. However, we include the newer data  at very high energies from the Large Hadron Collider (LHC) \cite{LHCtot1,LHCtot2,LHCtot3,totem2013,totem2013_2} and the Auger \cite{POAp-air} and HiRes \cite{HiRes} collaborations.

The fits were performed using the sieve algorithm \cite{sieve} to eliminate 13 outlying points among 179  total datum points.   Nine parameters were used in the fit leaving 157 degees of freedom, a total $\chi^2$ of 173.0, and a raw $\chi^2/{\rm d.o.f.} =  1.10$. This must be renormalized by the sieve factor ${\cal R}\approx 1.1$ to ${\cal R}\chi^2/{\rm d.o.f.}=1.21$ to account for the elimination of the outliers \cite{sieve}. The total $\chi^2$ would increase by 113.6 if we included the outliers, so the change would be substantial.  For comparison, the $\chi^2/{\rm d.o.f.}$ given by the fit is just 1.15 for  the $pp$ and $\bar{p}p$ total cross sections and $\rho$ values alone; much of the increase in the final result comes from the fit to the rather scattered values of $B$.  We note that {\em all} datum points including the outliers omitted in the final fit are shown in the figures comparing the fits with  data.

\begin{figure}[htbp]
\includegraphics{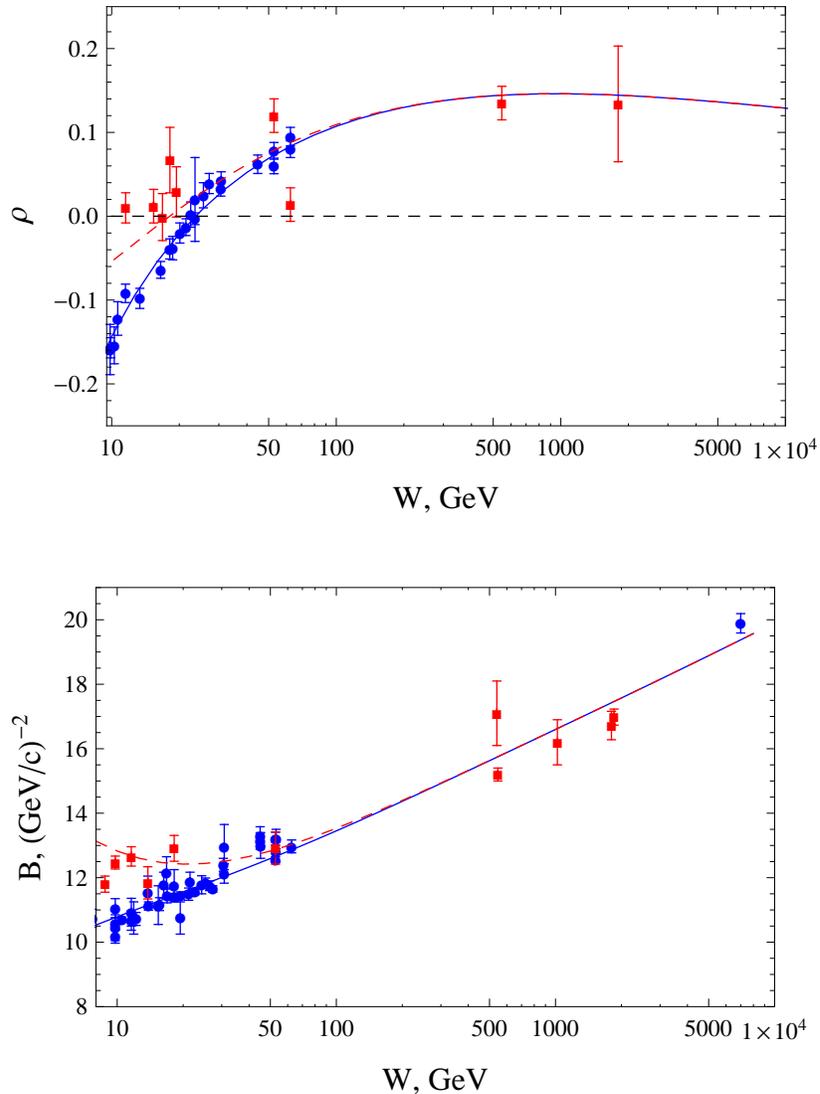}
 \caption{Top panel: fits to the ratios $\rho$ of the real to the imaginary parts of the forward scattering amplitudes for $pp$ (blue dots and solid line) and $\bar{p}p$ (red squares and dashed line) scattering. The horizontal dashed line is at $\rho=0$. Bottom panel: fits to the logarithmic slopes of the elastic differential scattering cross sections $d\sigma/dt$ for $pp$ (blue dots and solid line) and $\bar{p}p$ (red squares and dashed line) scattering. }
   \label{fig:rhoBfits}
\end{figure}

The results for the fits to the total and elastic scattering cross sections are shown in \fig{fig:crosssectionfits}.
The fits to the $\rho$ values and the logarithmic slopes $B$ of the forward differential elastic scattering cross sections $d\sigma/dt$, \eq{Bdef}, are shown in \fig{fig:rhoBfits}. The highest energy data for $\rho$ are from the LHC at 1,800 GeV. The value predicted for the LHC at $W=7$ TeV is $\rho=0.133$. The data for $B$ include the TOTEM results \cite{totem2013,totem2013_2} from the LHC at  $W=7$ TeV.

%

\begin{figure}[htbp]
\includegraphics{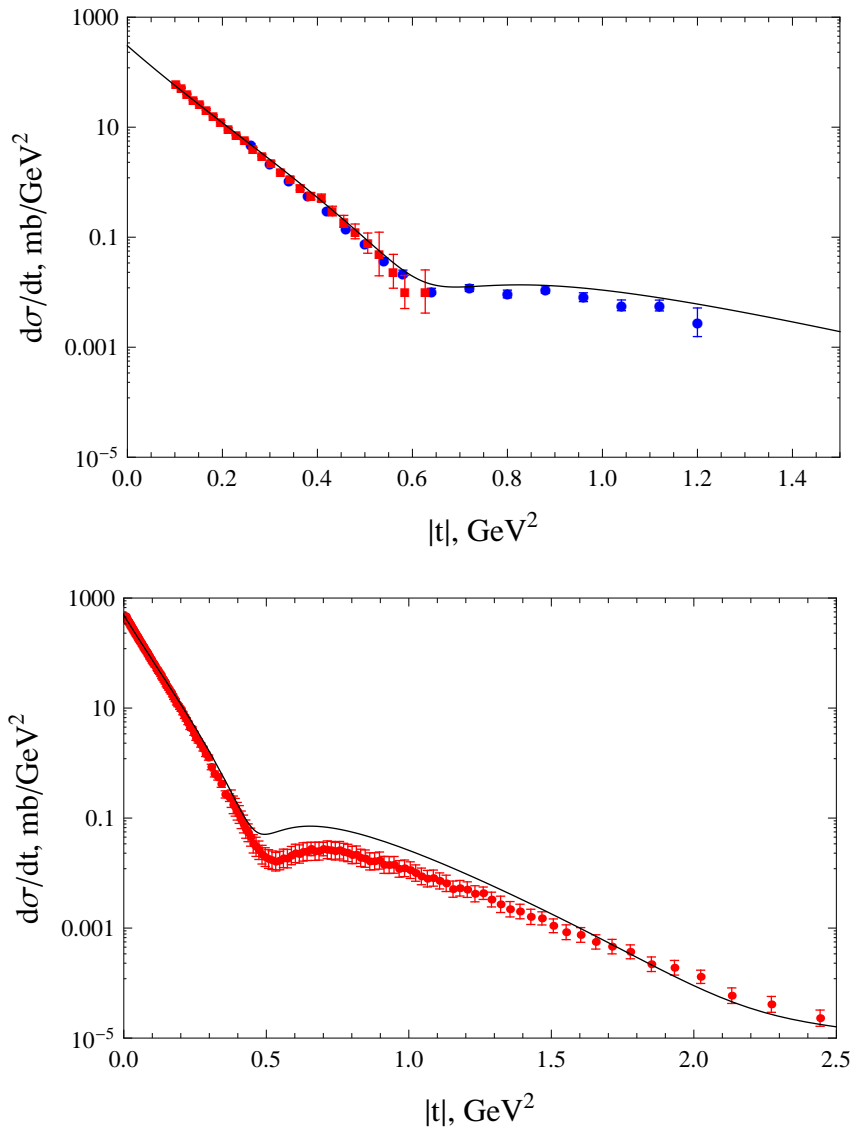}
 \caption{Top: the differential cross section $d\sigma/dt$ from the E710 experiment \cite{E710_B,E710_B2} at $W=1800$ GeV. Bottom $d\sigma/dt$ from the TOTEM experiment \cite{totem2011} at $W=7000$ GeV.}
 \label{fig:dsigdt}
 \end{figure}
\begin{figure}[htbp]
\includegraphics{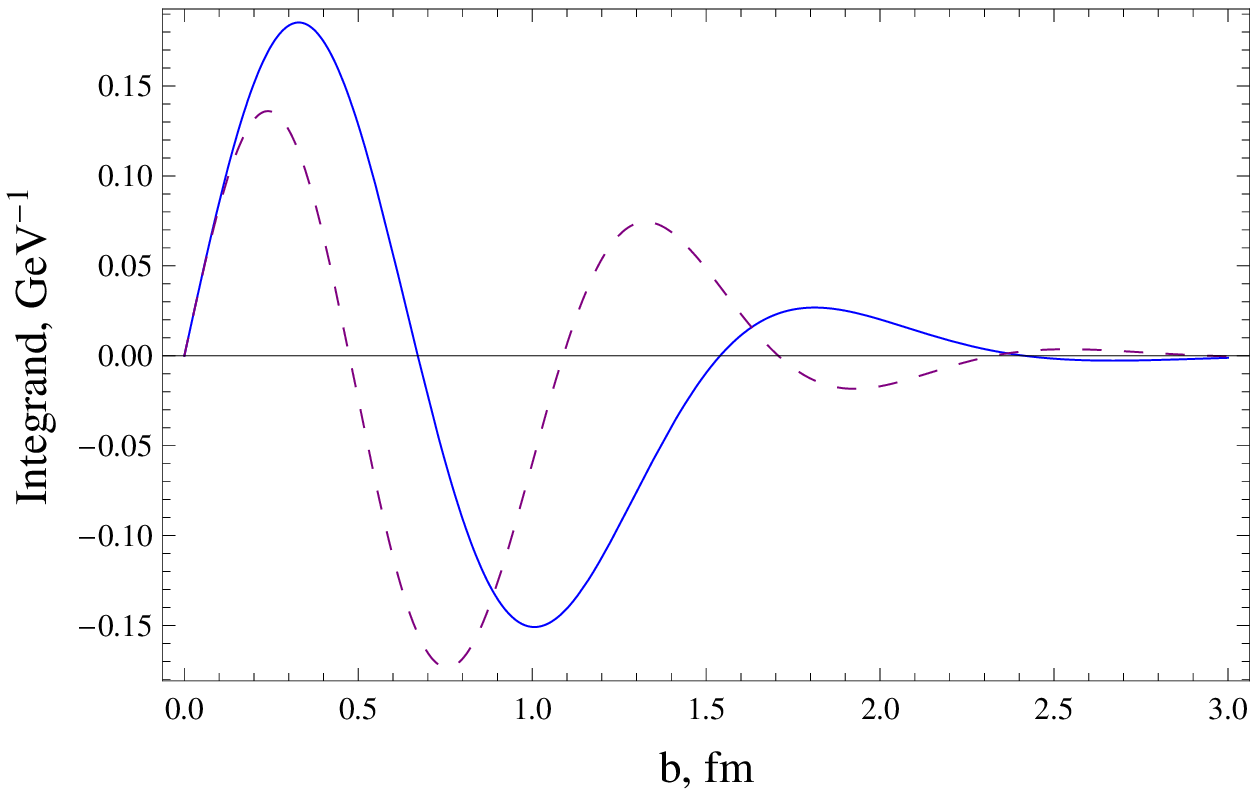}
 \caption{Plot of the integrand $b(1-\cos{\chi_R}e^{-\chi_I})J_0(b\sqrt{-t})$ for the imaginary part of $f(s,t)$ versus the impact parameter $b$ for $W=1$ TeV and $|t|=0.5$ GeV$^2$ (solid blue curve) and 1 GeV$^2$ (dashed purple curve).}
 \label{fig:Imfintegrand}
 \end{figure}

The measured and predicted differential cross sections $d\sigma/dt$ are shown in \fig{fig:dsigdt}  at $W=1800$ and 7000 GeV. Our descriptions of the cross sections at small $|t|$ are good, corresponding to our fits to the $B$ parameters, and the locations of the diffraction minima are reproduced properly. We are not concerned about the failure of our simple eikonal model to reproduce the differential cross sections in detail at large values of $|t|$ since the scattering amplitudes in this region are very sensitive to the cancellations which result from the oscillations of the Bessel function in \eq{f}, with the resulting scattering amplitudes of order $\sim 10^{-2} \times f(s,0)$.  As an illustration, we show the integrands for ${\rm Im}\,f(s,t)$ for $\sqrt{s}=W=1$ TeV and $|t|=0.5$ GeV$^2$ and 1 GeV$^2$ in \fig{fig:Imfintegrand}. The existence of large cancellations and the resulting sensitivity of the integrals to small details of the eikonal function not modeled here are evident. We emphasize, however, that the cross sections, $\rho$, and $B$ are much less sensitive to such details.

 %

 %


\section{Structure of the eikonal amplitudes and the edge in the $pp$ and $\bar{p}p$ scattering amplitudes \label{sec:structure}}

\subsection{Eikonal structure \label{subsec:structure}}

\begin{figure}[htbp]
\includegraphics{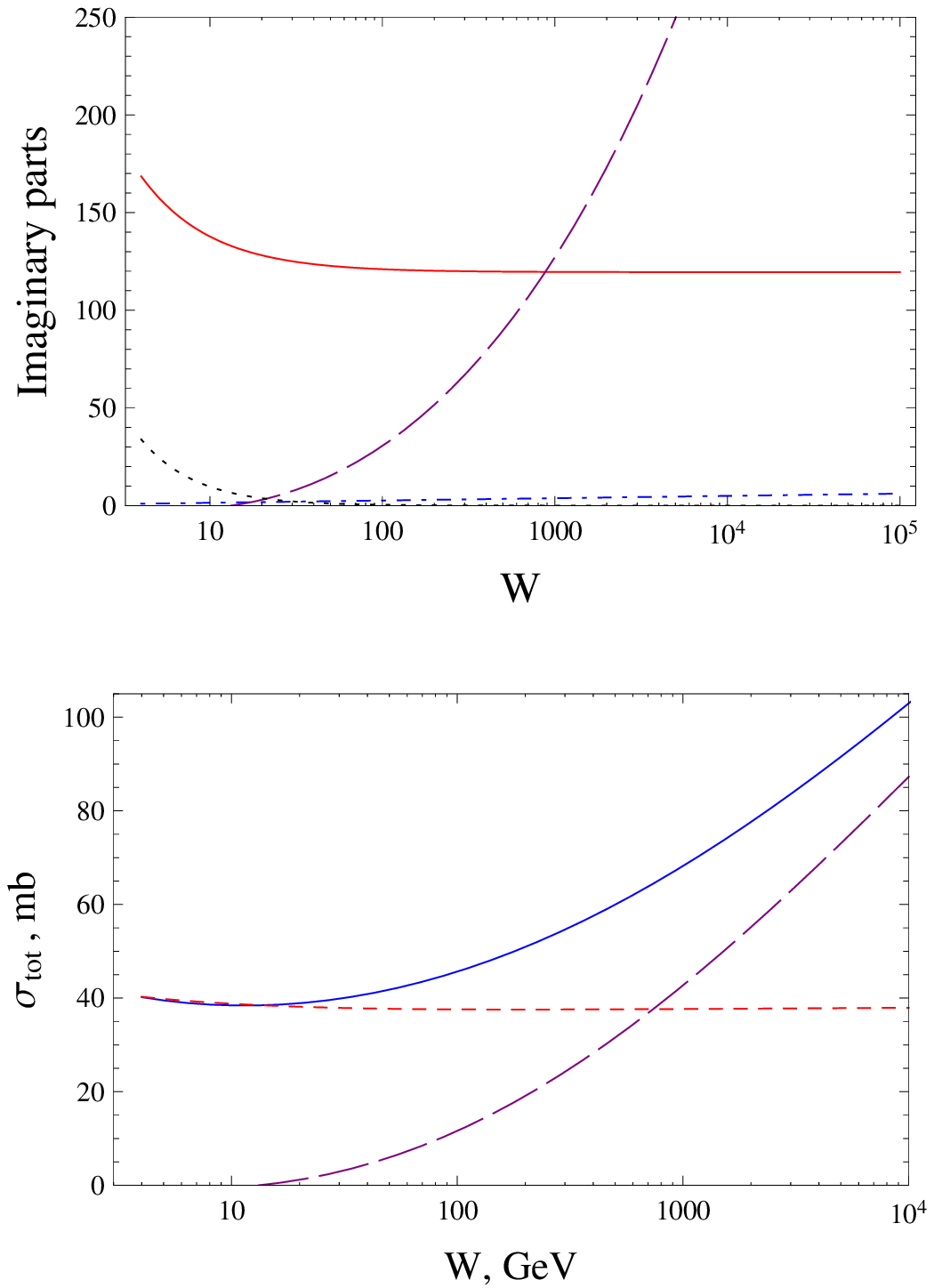}
 \caption{Top: Comparison of the imaginary parts of the different energy-dependent factors in the eikonal function. Solid (red) curve: $\sigma_{qq}$. Dot-dashed (blue) curve barely visible near zero amplitude: $\sigma_{qg}$. Long-dashed (purple) curve: $\sigma_{gg}$. Dotted (black) curve: the odd term.  Bottom: comparison of the cross sections calculated with (solid blue curve) and without ( dashed red curve) the inclusion of the gluon-gluon ($gg$) term in the eikonal function. The cross section for pure gluon scattering is shown as the long dashed purple curve. }
 \label{fig:sigcontributions}
 \end{figure}

It will be important for later interpretation to understand the relative importance of the various contributions to the eikonal function and cross sections. Since $\chi_R$ is small, $\chi_I$ determines the cross sections to good approximation. In the top panel of \fig{fig:sigcontributions} we therefore compare the imaginary parts of the energy-dependent  factors in $\chi_I$. The most important contribution at high energies ({\em e.g.}, $W\gtrsim 1$ TeV) is clearly that from gluon-gluon scattering, parametrized in our model in a form suggested  by the rapid growth of $gg$ scattering in perturbative QCD and the corresponding minijet models for the rise in $\sigma_{\rm tot}$ with energy.

 %

The other important contribution at high energies is that labelled $\sigma_{qq}$. This term cannot be separated in the fit from $\sigma_{qg}$, and is not to be interpreted strictly in terms of quark-quark scattering in the sense of the parton model. The $qq$ scattering in that model in fact becomes small at lower energies, while the combined contributions of  $\sigma_{qq}$ and $\sigma_{qg}$  increase as parametrized. The low-energy behavior presumably arises from ``soft'' processes such as the scattering of valence quarks evident, for example, in the approximate 2/3 ratio of the $\pi p$ and $pp$ cross sections, and to Regge exchange terms.

The crossing-odd contribution $\chi_O$ vanishes rapidly with increasing energy. Finally, the mixed quark-gluon term, parametrized in a form suggested by minijet models \cite{blockaspen}, is strongly mixed and correlated with the other terms in the fitting, and should not be interpreted directly in terms of $qg$ scattering.

\begin{figure}[htbp]
\includegraphics{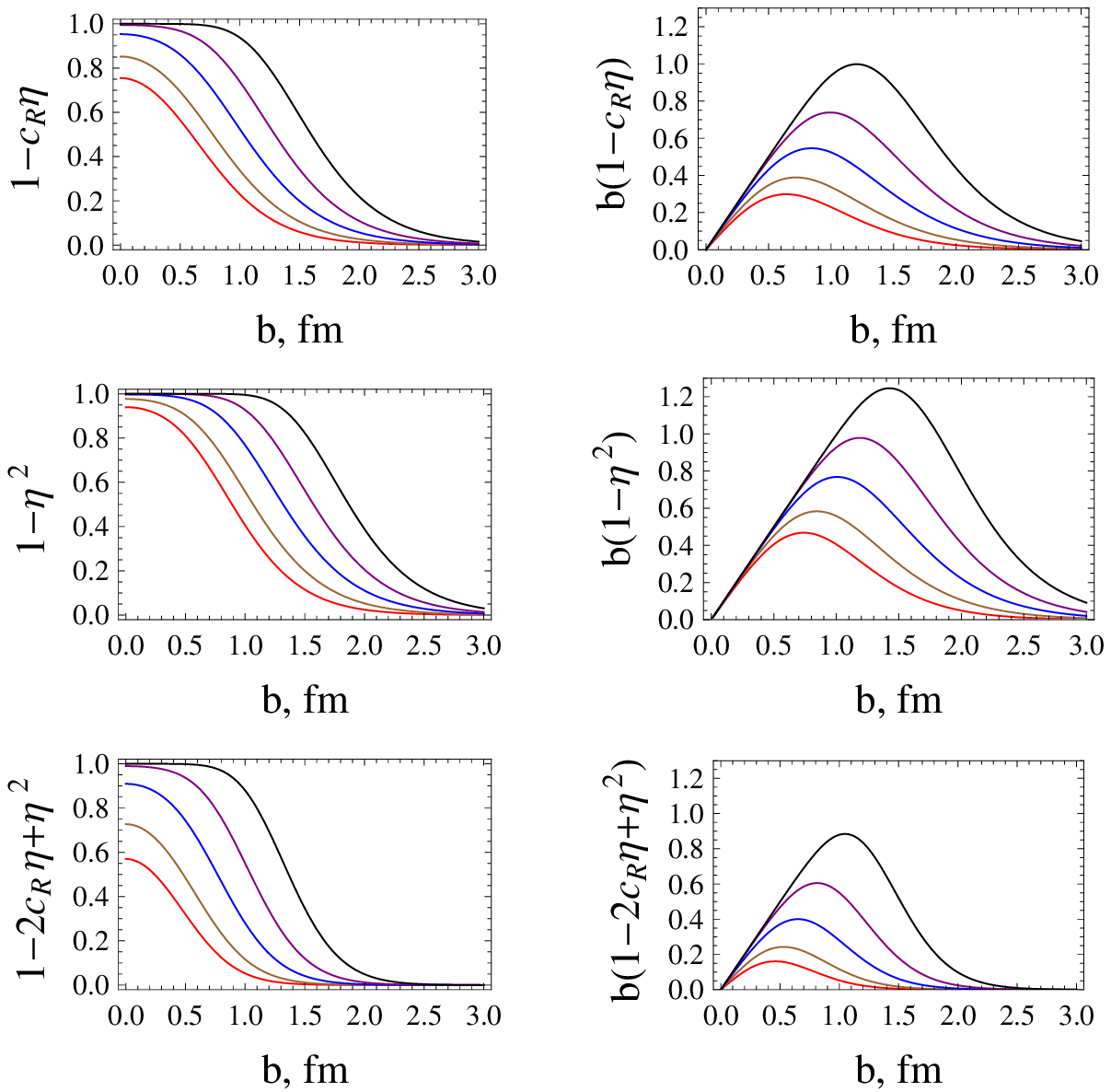}
 \caption{Plots of the eikonal factors (left-hand column) and those factors multiplied by the geometric factor $b$ in the integrands for the $pp$ total cross section (top row),  inelastic cross section (middle row), and elastic cross section (bottom row). In the labels for the ordinate, $\eta=e^{-\chi_I}$ and $c_R=\cos{\chi_R}$. The curves in each panel correspond, bottom to top, to energies $W=50$ GeV (red curve), 500 GeV (brown curve), 5 TeV (blue curve), 50 TeV (purple curve) and 1000 TeV (black curve). }
 \label{fig:ppintegrands}
 \end{figure}

The eikonal factors and the complete integrands in the expressions for $\sigma_{\rm tot}$, $\sigma_{\rm inel}$, and $\sigma_{\rm elas}$  in Eqs.\ (\ref{sigma_el})-(\ref{sigma_inel}) are shown for $pp$ scattering at a progression of energies in \fig{fig:ppintegrands}. The quantities $\eta$ and $c_R$ used in the labels in this figure are $\eta=e^{-\chi_I}$ and $c_R=\cos{\chi_R}$; similarly, $s_R=\sin{\chi_R}$.
%
 %

We note several important features of the curves shown. First,  the eikonal factors $(1-c_R\eta)$ for $\sigma_{\rm tot}$ and  $(1-2c_R\eta+\eta^2)$ for $\sigma_{\rm elas}$
lie well below the asymptotic ``black disk'' limit 1 at small values of the impact parameter $b$ for energies $W\lesssim 5$ TeV  as shown in the left-hand column in \fig{fig:ppintegrands}. The scattering is far from asymptotic, and the approach to an asymptotic distribution flat at the value 1 out to a sharp cutoff radius $\sim R$ is extremely slow. The inelastic integrand approaches asymptotic-like behavior more rapidly, with the eikonal factor  $1-\eta^2\approx 1$ becoming flat at 1  at small $b$, at a sharper cutoff, at lower energies.

The actual integrands including the geometric factor $b$ are shown  in the right-hand column in \fig{fig:ppintegrands}. This factor pushes the relevant impact parameters toward larger $b$ and introduces the peaked behavior shown. The main integrals involved in the calculation of the logarithmic slope parameter $B=d[\ln{(d\sigma/dt)}]/dt|_{t=0}$, \eq{B} or (\ref{Bapprox}), involve an extra factor $b^2$ in the numerator, with the result that the main contributions to the integral are pushed to  larger values of $b$ and become increasingly sensitive to the tail of the eikonal distribution for $\sigma_{\rm elas}$.

The integrand for $\rho$, the ratio of the real to the imaginary parts of the forward elastic scattering amplitude, \eq{rho}, involves a factor $\eta=e^{-\chi_I}\sin{\chi_R}$ in the numerator. Since the transparency factor $\eta$ vanishes strongly at small $b$ at high energies, the result of strong inelastic absorption, the main contributions to $\rho$ are pushed toward higher values of $b$, beyond the peak in the integrand for $\sigma_{\rm elas}$, and are again more sensitive to the tail of the distribution than the elastic scattering cross section itself. In addition, the phase of the scattering amplitude, hence the ratio of $\chi_R$ to $\chi_I$, is determined at high energies by the form assumed for $\chi$ coupled with the constraints imposed by the analyticity of the scattering amplitude \cite{blockcahn,blockrev}, $\chi_R$ is not freely variable. As a result, there is a tension between $B$ and $\rho$ when fitting data: both are sensitive to the tail of the distribution and the two parameters are therefore coupled with respect to changes in that distribution. The fit to the current data discussed above incorporates the constraints imposed by both $B$ and $\rho$.


\subsection{Connection  to the description in terms of real analytic ampitudes \label{subsec:BHconnection}}

Despite the seeming complication of our fit to the data in terms of the eikonal function, the results we obtain for $\sigma_{\rm tot}$ and $\rho$ for $pp$ and $\bar{p}p$ scattering can be described very well with an expression of the simple form used by Block and Halzen \cite{blockhalzenfit,blockrev} in their earlier fit to the corresponding data up to $W=1800$ GeV. That fit gave successful predictions of the more recent, higher energy data \cite{blockhalzen,blockhalzen2}.

The Block-Halzen analysis assumed a $\ln^2s$ bound on the growth of the cross sections at high energy, imposed the constraints implied by the analyticity of the scattering amplitudes, and was constrained to connect smoothly to the low-energy data.
It was based on the use of analytic amplitudes of the form \cite{blockrev}
\ba
\label{BHsigmas}
\sigma^{\pm}(\nu) &=& c_0+c_1\ln\left(\frac{\nu}{m}\right)+c_2\ln^2\left(\frac{\nu}{m}\right)+\beta'\left(\frac{\nu}{m}\right)^{\mu-1} \pm\delta \left(\frac{\nu}{m}\right)^{\alpha-1}, \\
\label{BHrhos}
\rho^{\pm} &=& \frac{1}{\sigma^{\pm}}\left\{\frac{\pi}{2}c_1+\pi c_2\ln\left(\frac{\nu}{m}\right) - \beta' \cot\left(\frac{\pi\mu}{2}\right) \left(\frac{\nu}{m}\right)^{\mu-1} +\frac{4\pi}{\nu}f_+(0)
 \pm \delta\tan\left(\frac{\pi\alpha}{2}\right)\left(\frac{\nu}{m}\right)^{\alpha-1}\right\},
\ea
where the upper and lower signs are for $pp$ and $\bar{p}p$ scattering respectively. Here $\nu$ is the laboratory energy of the incident particle, with $2m\nu=s-2m^2=W^2-2m^2$ where $m$ is the proton mass.

Their fit used the then-extant data on $\sigma_{\rm tot}$ and $\rho$ in the range $6\leq W \leq 1800$ GeV plus analyticity constraints on the values and slopes of the cross sections at $W=4$ GeV which followed from finite-energy sum rules applied to the data at lower energies. The fit was excellent and led to successful predictions of the results for the cross sections later measured at the LHC and in cosmic ray experiments \cite{blockhalzen,blockhalzen2}. A later analysis of the inelastic scattering data using an expression of the same form as \eq{BHsigmas} was also successful and gave evidence of an approach to the black disk limit at ultrahigh energies \cite{blockhalzen}, with $\sigma_{\rm inel}\rightarrow \sigma_{\rm tot}/2$.

 We have checked that the use of the Block-Halzen expressions to fit ``data'' derived from our results gives curves for $\sigma_{\rm tot}$ and $\sigma_{\rm elas}$ that are almost indistinguishable from the curves in \fig{fig:crosssectionfits}. Both fits describe the data quite well, and we conclude that they are consistent.

The expressions in \eq{BHsigmas} and \eq{BHrhos} simplify to a more familiar form for $W^2\gg m^2$, with $\nu/m\rightarrow W^2/2m^2=s/2m^2$. The corrections to the logarithmic terms are negligibly small for $W$ in the region of the fit. The corrections to the power-law terms are a fraction of a millibarn for $W=4$ GeV, negligible for $W\gtrsim 6$ GeV, and can be absorbed overall in slight adjustments of the powers and coefficients of those terms where they are relevant. As a result, the argument $\nu$ in the formulas for $\sigma^{\pm}(\nu)$ and $\rho^{\pm}(\nu)$ can be converted directly to $s/2m^2$, or with some rearrangement of terms and coefficients, to $W/m$, without loss of accuracy in the region used in our fit and that of Block and Halzen.  A similar expansion quadratic in $\ln(s/m^2)$  for $\sigma_{\rm elas}$ follows from the results in \cite{blockhalzen}.    The coefficient of $\ln^2(s/m^2)$ in $\sigma_{\rm inel}$ was  found in \cite{blockhalzen} to be one-half that found for $\sigma_{\rm tot}$  as required for an asymptotic black-disk limit for the scattering amplitude.

 A quadratic in $\ln{s}$ was also used by Schegelsky and Ryskin \cite{ryskin} to fit the data on $B$. Their result for the coefficient of $\ln^2s$ was consistent with the Block-Halzen fit to the cross sections and $\rho$ alone and the expectation for black disk scattering that $B\rightarrow R_{\rm tot}^2/4=\sigma_{\rm tot} /8\pi$ for $s\rightarrow\infty$.

Similarly, our results for  $B_{pp}$ can be written to an accuracy of a few parts in 1000 in the same form as
\be
\label{Bppseries}
B_{pp}(W) = 7.229 + 1.0862\ln(W/m_0) + 0.02209 \ln^2(W/m_0) + 3.719 (m_0^2/W^2)^{1/2}\ \  {\rm GeV}^{-2}
\ee
for $6\ {\rm GeV}\leq W \leq 10^{10}\ {\rm GeV}$, where $m_0=0.6$ GeV is our scale factor and $W=\sqrt{s}$. This result is potentially useful  in the conversion of cosmic ray cross sections for proton-air scattering to $pp$ cross sections.

While our eikonal fit to the complete data set gives results for the cross sections and $\rho$ that are essentially equivalent numerically to those of Block and Halzen over the energy range currently accessible, it is not immediately clear analytically from the rather complicated eikonal expressions  why the simple expressions in Eqs.\ (\ref{BHsigmas}) and (\ref{BHrhos}) --- or their reduced high-energy forms --- should work so well.  However,  the asymptotic $\ln^2s$ growth of the cross sections in \eq{BHsigmas} and the approach to the black-disk limit of the scattering follow directly from the expected power-law  growth of the eikonal function with $s$ coupled with the exponential cutoff in the overlap functions $A(b,\lambda)$, \eq{overlap}, for $\lambda b\gg 1$, an argument  familiar in discussions of the Froissart bound. The first  leads to strong growth of $\chi_I(b,W)$ with $\chi_I\gg1$ and $e^{-\chi_I}\ll 1$  at large $s$ and small $b$,  and a corresponding saturation of the scattering amplitudes for $\sigma_{\rm tot}$, $\sigma_{\rm inel}$, and $\sigma_{\rm elas}$ at the value 1 as seen in  the left-hand column in \fig{fig:ppintegrands}.  This saturation persists out to values of $b$ such the exponential decrease in $A(b,\lambda)$ pushes $\chi_I$ to values below 1 beyond which the scattering amplitudes vanish exponentially. The rough condition $\chi_I(b,W)\lesssim 1/2$ determines the effective cutoff radius in $b$ which, given the exponential behavior of $A(b,\lambda)$, can grow only as $\ln{s}$.

Simple arguments using Eqs.\ (\ref{sigma_el})--(\ref{sigma_inel}) and (\ref{B}) or (\ref{Bapprox})  then show that $\sigma_{\rm elas}$ and $\sigma_{\rm inel}$ tend to $\sigma_{\rm tot}/2$  for $s$ sufficiently large, while $B\rightarrow \sigma_{\rm tot}/8\pi$, with all proportional to $\ln^2s$ plus logarithmic and lower-order corrections associated with the edge region in the amplitudes, the region around the peaks in the integrands shown in the right-hand column of \fig{fig:ppintegrands}.  These arguments provide a justification for the Block-Halzen form for the cross sections at sufficiently high energies; detailed checks using our eikonal fit show that the simple quadratic expressions are accurate at present-day energies. Finally, one can show from \eq{rho} and the constraint on the phase of the scattering amplitude imposed by analyticity and unitarity \cite{blockcahn,blockrev} that $\rho\rightarrow 0$ at very high energies.

We emphasize that the eikonal fit allows us to calculate important quantities such as $B$ and the differential scattering cross sections $d\sigma/dt$ that are not accessible through a Block-Halzen type analysis without further input. Our analysis of the eikonal results on $B$ shows that it, like the cross sections, can be described to high accuracy by a quadratic in $\ln{s}$ plus low-energy Regge-like terms, thus providing the necessary input.  Given the uncertainty in the rate of the power-law growth of the eikonal function, and the uncertainty in its form at large $b$ discussed in Sec.\ \ref{subsec:edge_origin},  we believe that fits to the data using the quasi-universal high-energy expressions in Eqs.\ (\ref{BHsigmas}) and (\ref{BHrhos}), the corresponding result for $\sigma_{\rm inel}$ in \cite{blockhalzen}, and the expansion above for $B$, are likely to give a more reliable way  at this point of extrapolating the cross sections to ultrahigh energies .


\subsection{The edge of the $pp$ and $\bar{p}p$ scattering amplitudes \label{subsec:edge}}

Block {\em et al.} \cite{edge} recently established that the proton-proton  scattering amplitude in impact parameter space has an edge region the width of which is essentially constant over many orders of magnitude in the center-of-mass energy $W$. This result followed from the usual form for the scattering amplitude combined with the assumption that the scattering is strongly absorptive.  In particular, it was shown in \cite{edge} that this edge could be isolated using the properties of the transparency function $\eta(b,s)\equiv \exp{[-\chi_I(b,s)]}$. This function is very small at small impact parameters where the absorption is strong, then rises to unity --- complete transparency and no scattering --- at large impact parameters.

 This observation was exploited in \cite{edge} by noting that
\ba
\label{edgeintegral}
\sigma_{\rm tot}-2\sigma_{\rm elas} &=& 2\sigma_{\rm inel}-\sigma_{\rm tot}\\
\label{edgeintegral2}
& = & 4\pi\int_0^\infty db\,b\,\eta(\cos{\chi_R}-\eta)  \\
&\approx&  4\pi\int_0^\infty db\,b\,\eta(1-\eta)
\label{edgeintegral3}
\ea
for small real parts of the scattering amplitude, a condition satisfied in the present fit. The integrands in these expressions have the property that they are large only in the transition region between strong absorption and no scattering.

Since $\sigma_{\rm tot}$ , $\sigma_{\rm elas}$, and $\sigma_{\rm inel}$ are measured quantities, experiment gives a direct measurement of the edge integral in \eq{edgeintegral}. To obtain its extrapolation to very high energies, Block {\em et al.} \cite{edge} used the very accurate Block-Halzen fit \cite{blockhalzenfit,blockrev} to the $pp$ and $\bar{p}p$ total cross sections and $\rho$ values for $1800\geq W\geq 6$ GeV, and its extension to $\sigma_{\rm inel}$  \cite{blockhalzen}. This fit, which  incorporated the asymptotic $\ln^2s$ limit on the growth of the total cross sections for large $s=W^2$ and the constraints imposed by the analyticity of the scattering amplitudes and the lower energy data, successfully predicted the recent LHC and cosmic ray results \cite{blockhalzen,blockhalzen2}. The constancy of the edge width followed directly from the use of those results in \eq{edgeintegral}, and did not depend on the detailed impact parameter distribution in \eq{edgeintegral3}.

\begin{figure}[htbp]
\includegraphics{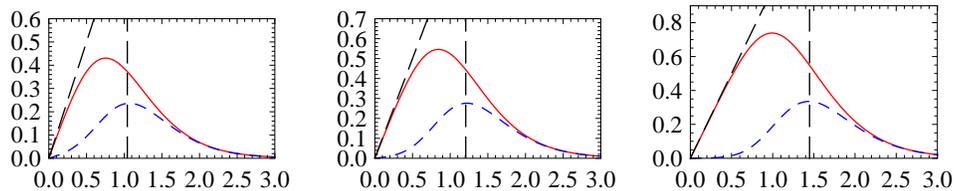}
  \caption{Comparisons of the integrands $b(1-\cos{\chi_R}e^{-\chi_I})$ for $\sigma_{\rm tot}$ (solid red curve), the ``black disk'' integrands for disk radius $R_{\rm tot}=\sqrt{\sigma_{\rm tot}/2\pi}$ (long-dashed black curve), and  the edge integrands $b(\cos{\chi_R}-e^{-\chi_I})e^{-\chi_I}$ (short-dashed blue curves) at energies $W=1,\,5,$ and 50 TeV, left to right. The vertical and horizontal scales give the integrand and $b$ in fm.}
  \label{fig:sigmaedgeRcomp}
 \end{figure}

In the present eikonal fit to the $pp$ and $\bar{p}p$ data, the edge integrand $b\eta(\cos{\chi_R}-\eta)\approx b\eta(1-\eta)$ is peaked at values of the impact parameter somewhat beyond the peak in the integrand for $\sigma_{\rm tot}$ as shown in \fig{fig:sigmaedgeRcomp} and well into the tail region in the eikonal distribution for $\sigma_{\rm tot}$ as can be seen by a comparison to \fig{fig:ppintegrands}, top left. Not surprisingly, this is just the region that determines the effective black disk radius $R_{\rm tot}=\sqrt{\sigma_{\rm tot}/2\pi}$ of the scattering amplitude. As seen in the comparison of the actual $pp$ scattering amplitude with the black disk amplitude  with the same value of $\sigma_{\rm tot}$ in \fig{fig:sigmaedgeRcomp},  the ``missing'' contributions to the black disk amplitude for $b<R_{\rm tot}$ are supplied by the tail of the distribution with $b>R_{\rm tot}$, with $R_{\rm tot}$ corresponding very closely to the peak in the edge integrand.

 %

The value of the edge  integral should be approximately the height of the peaked integrand times its width  $t_{\rm edge}$ at half maximum. We define $t_{\rm edge}$ as  the edge width. Since $\eta(1-\eta)$ has a maximum value of 1/4 and $b$ is approximately equal to $R_{\rm tot}$ at the peak,
\be
\label{edge_sigma}
\sigma_{\rm tot}-2\sigma_{\rm elas}\approx \pi R_{\rm tot} t_{\rm edge},
\ee
 or
\be
\label{edgewidth_t}
t_{\rm edge} \approx (\sigma_{\rm tot}-2\sigma_{\rm elas})/\sqrt{\pi\sigma_{\rm tot}/2}.
\ee

\begin{figure}[htbp]
\includegraphics{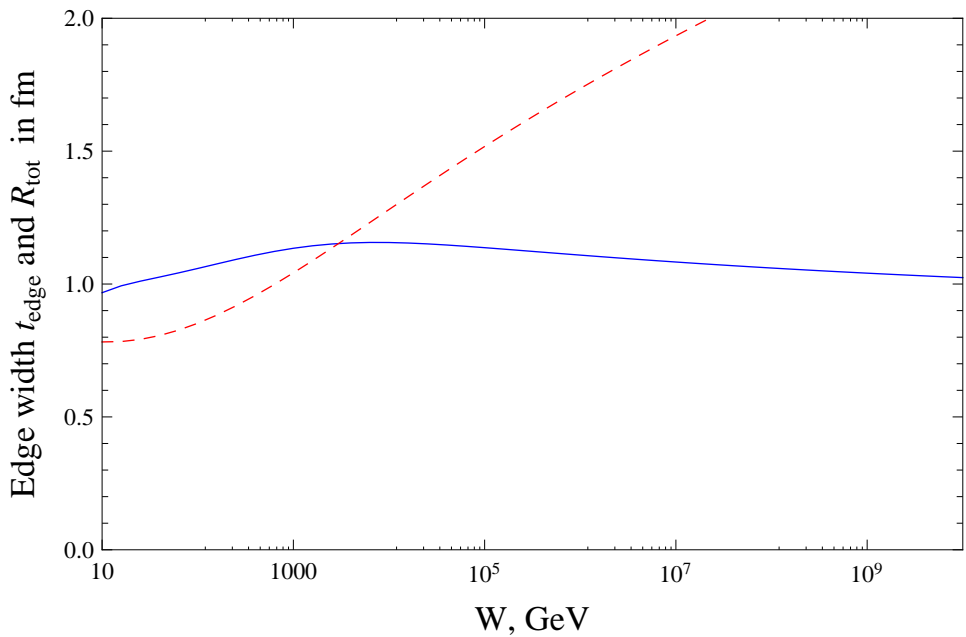}
 \caption{Plots of the $pp$ edge width $t_{\rm edge}$ calculated using the present eikonal fit to the $pp$ and $\bar{p}p$ data (solid blue curve)
  and the  black disk radius $R_{\rm tot}=\sqrt{\sigma_{\rm tot}/2\pi}$ (dashed red curve) as functions of the center-of-mass energy $W$.}
 \label{fig:edgewidthRcomp}
 \end{figure}

The edge width $t_{\rm edge}$ was evaluated in \cite{edge} using the Block-Halzen fit to the $pp$ and $\bar{p}p$ total cross sections and $\rho$ values \cite{blockrev} and its extension to the inelastic cross sections \cite{blockhalzen}. It was found to be remarkably constant at $t_{\rm edge}\approx 1$ fm above  about 10 GeV. The result obtained here using our fit to the $pp$ and $\bar{p}p$ data is essentially the same; this as shown in \fig{fig:edgewidthRcomp}.

 %

\begin{figure}[htbp]
\includegraphics{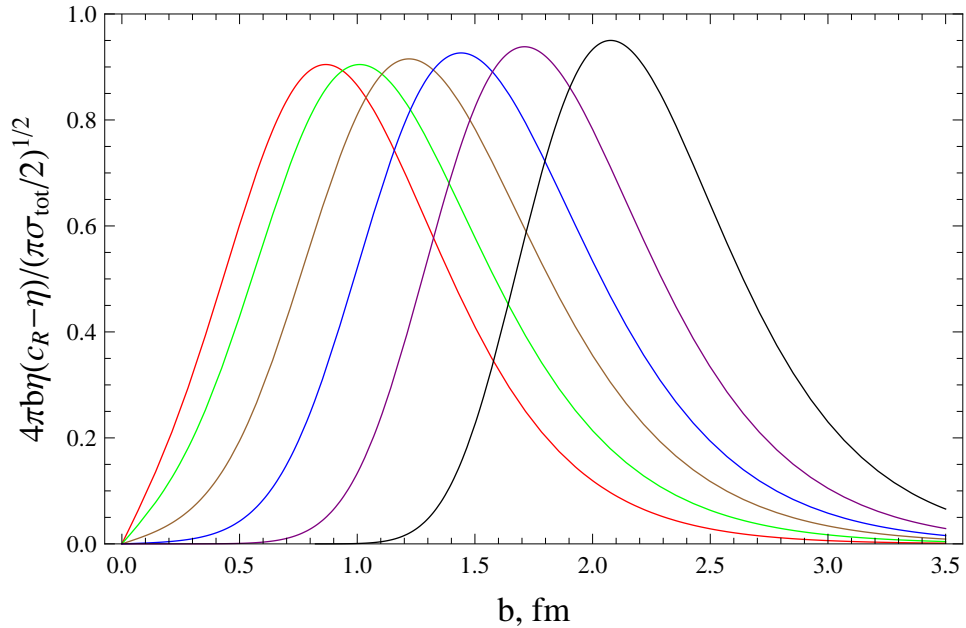}
 \caption{Plots of the normalized edge integrands $b(\cos{\chi_R}-e^{-\chi_I})e^{-\chi_I}/R_{\rm tot}$ for, left to right, $W=30$ GeV (red), 500 GeV (green), 5000 GeV (brown), 5$\times 10^4$ GeV (blue), $10^6$ GeV (purple), and $10^8$ GeV (black).}
 \label{fig:edgeplotnorm}
 \end{figure}

The reason for this result can be seen in \fig{fig:edgeplotnorm} where we plot the edge integrand normalized to $R_{\rm tot}$. The resulting distributions have an approximately fixed shapes and areas as functions of $W$, and migrate slowly to larger values of $b$ with increasing energy. We emphasize that these properties are determined up to 7 TeV by our fit to data on $\sigma_{\rm tot}$ and $\sigma_{\rm inel}$ or $\sigma_{\rm elas}$. The fits to $\sigma_{\rm tot}$ extend to $W\sim70$ TeV. The results shown in \fig{fig:edgewidthRcomp} at higher energies give our predictions based on the present eikonal model; the results are consistent with those of Block {\em et al.} \cite{edge} which are independent of a detailed eikonal description of the scattering.

 %

 As seen in \fig{fig:edgewidthRcomp}, $t_{\rm edge}$ and $R_{\rm tot}$ cross in magnitude for $W\approx 2$ TeV, with $R_{\rm tot}$ larger and increasing at higher energies. The eikonal amplitude for $\sigma_{\rm tot}$ is also beginning to saturate at 1 at small $b$ in this region. The crossover  point therefore gives a reasonable estimate of the energy at which the scattering amplitude begins to show aspects of asymptotic behavior, with the edge region becoming less important than the central region. We note that the crossover region is where the gluon contributions to the cross section become dominant as is evident in \fig{fig:sigcontributions}.


\subsection{Origin of the edge \label{subsec:edge_origin}}

The two key results discussed above are: (1), the logarithmic growth with energy of the effective radius  of the strongly absorptive black disk region in the scattering amplitude and the resulting $\ln^2s$ growth of $\sigma_{\rm tot}$;  and, (2), the existence of an edge region with constant width $t_{\rm edge}$.

There are several possible explanations for the origin and constancy of the edge in the $pp$ and $\bar{p}p$ scattering amplitudes. A classic explanation would attribute the edge to pionic fluctuations around the proton or antiproton, with the pionic fluctuations then interacting in the collision. This would give an edge region on the scale of $1/m_\pi\approx 1.4$ fm, but does explain how the overall radial scale of the scattering amplitude would increase, or how strongly the fluctuations would couple to the the expanded proton in this picture.

The most likely {\em combined} explanation for both effects in our view is connected to the rapid increase in the strength of the gluon fields in the interacting hadrons with increasing energy, and the resulting saturation of the exponentially bounded gluon confinement volumes of the two particles. This leads to a growing likelihood of interaction in a $pp$ or $\bar{p}p$ collision and  gluon dominance of the scattering, with the radius of the region of strong absorption growing as $\ln{s}$.   Specific models based on parton collisions \cite{fagundes} and Reggeon field theory \cite{KhozeMartinRyskin,GotsmanLHC} have this character.

This picture should hold in any hadronic scattering at energies sufficiently high  that the quark effects prominent at low energies can be neglected. Since the gluon confinement volume is presumably fixed, all hadronic cross sections should then have a universal ${\rm constant}\times\ln^2s$ behavior at high energies, with a common value of the multiplicative constant. This behavior is consistent with the observed behavior of all cross sections which are known at high energies \cite{PDG2014}.

The scale of the gluon confinement volume is set in our model by $1/\mu_{gg}$. The increasing strength of the gluon fields is represented in QCD-based minijet-type models by the growth of the gluon distribution functions $f_g$ with decreasing Bjorken $x$ values or increasing energy of the gluon-gluon collision. This leads to stronger gluon-gluon scattering in the collision, typically increasing as a power of $s$, corresponding growth of $\chi_I$, and gluon dominance of high-energy scattering.

Our eikonal model and that of Fagundes {\em et al.\/} \cite{fagundes} have this general character. The power-law growth of $\sigma_{gg}$ with $s$, combined with the exponential cutoff in the overlap function $A(b,\mu_{gg})$ at large impact parameters, is sufficient to ensure an asymptotic $\ln{s}$ growth of $R_{\rm tot}$ and $\ln^2s$ growth of the cross sections at very high energies independent of any appeal to the Froissart bound  \cite{froissart,martin1} on the cross sections.\footnote{The Froissart bound is just that, a bound on the growth of $\sigma_{\rm tot}$ with $s$. The energy scale in the rigorous bound is set by the lightest mass scale in the scattering. This is not relevant at high energies where we expect the mass scale to be that of the confinement volume, here $\mu_{gg}$, outside of which the hadrons interact only rarely. Since $1/\mu_{gg}\ll 1/m_\pi$, the rigorous bound is far from being saturated.} The dominance of the gluons at high energies is evident in \fig{fig:sigcontributions}. In the black disk limit, $\sigma_{\rm elas},\ \sigma_{\rm inel}\rightarrow\sigma_{\rm tot}/2$, the leading $\ln^2s$ terms in the cross sections cancel in \eq{edgeintegral}. The asymptotic parametrizations in \eq{BHsigmas} and the equivalents for $\sigma_{\rm elas}$ and $\sigma_{\rm inel}$ combined with the logarithmic growth of $R_{\rm tot}$ then indicate that $t_{\rm edge}$ should be constant, or nearly so, at high energies as is observed.

The width of the edge is also related to $1/\mu_{gg}$ in our model: the factor $\eta$ in the edge integrand $\eta(1-\eta)$, \eq{edgeintegral} increases from 0 to 1 over an interval in $b$ proportional to $1/\mu_{gg}$ centered around $R_{\rm tot}$, while the factor $(1-\eta)$ falls from 1 to 0 over a similar interval. This results in the edge integrands shown in \fig{fig:edgeplotnorm} with the width of the peaks, hence $t_{\rm edge}$, proportional to $1/\mu_{gg}$.

A different, but potentially  related, mechanism was proposal by Rosner \cite{rosner}, who described the edge in terms of the breaking of flux strings connecting quarks or $3\bar{3}$ gluon configurations in the two hadrons. His estimate of the edge width, based on the energy needed to break such QCD strings in other processes, is of the right general size. We attribute the scattering mainly to interactions of the gluons or gluon fields in the overlapping nucleons in the collision. Some components of those fields may be mixed between the nucleons in the scattering, leading to the stretching of flux tubes between the nucleons as they separate and  extending the effective range of the interaction through string breaking as Rosner proposed. The growth in radius of the main interaction region, however,  arises from the increasing saturation of the gluon confinement volume as sketched above.

Since gluon dominance should appear at high energies in all hadron-hadron scattering,  we expect all hadron-hadron total cross sections to approach a universal $\ln^2s$ growth with a common coefficient at very high energies as noted above.  All hadron scattering amplitudes should also have an edge region  with an approximately constant  width proportional to $1/\mu_{gg}$: the leading terms in the cross section difference in \eq{edgewidth_t} and its analog for other hadrons cancel in the difference given their universal behavior at high energies, and  the subleading terms are presumably logarithmic in $s$ as is the factor in the denominator.

The asymptotic mass scale and the corresponding behavior of the eikonal function at large impact parameters  are clearly of considerable interest. The model we have used assumes that $\chi(b,s)$ can be written as a sum of terms in which the energy dependence factors out of the overlap functions $A(b,\lambda)$ where the latter, exploiting ideas originally formulated by Wu and Yang \cite{wuyang}, are given as  convolutions of density distributions similar to those associated with the proton charge and magnetic moment form factors. The resulting overlap functions are small at large impact parameters, so the integrand for the scattering amplitude $f(s,t)$ is  proportional   to $\sigma_{gg}(s)\times\mu_{gg}^2(\mu_{gg}b)^3K_3(\mu_{gg}b)$ at at large $b$ for energies where gluon scattering is dominant.

This is inconsistent on the surface with the result expected from the dispersion relation in $t$ for $f(s,t)$, schematically
\be
\label{dispersion}
f(s,t) = \int_{t_0}^\infty dt'\,\frac{a(s,t')}{t'-t}.
\ee
The partial wave amplitude for angular momentum $j$ is just
\be
\label{f_j}
f_j(s) =  \frac{1}{2p^2}\int_{t_0}^\infty dz\,a(s,t')Q_j\left(\frac{t'}{2p^2}+1\right).
\ee
Using the standard approximation $Q_j(z)\approx K_0\left(\sqrt{2j^2(z-1)}\right)$, very good for $j$ large and $(z-1)$ small, and introducing the impact parameter $b=j/p$, we get the impact parameter representation
\be
\label{f(b)}
f_j(s)\longrightarrow f(s,b)=\frac{1}{2p^2}\int_{t_0}^\infty dt'\,a(s,t') K_0\left(b\sqrt{t'}\right).
\ee

The functions $x^3K_3(x)$ and $K_0(x)$  behave quite differently for $x$ large, and \eq{f(b)} involves an integral over $t'$ while the $gg$ eikonal function involves only the fixed scale $\mu_{gg}$. There is consequently some uncertainty as to how well the asymptotic behavior of the scattering amplitude is described at large $b$ in the present model. Seen a different way, the weight function $a(s,t')$ for the $gg$ term is proportional to $\delta(t'-\mu_{gg}^2)/(t'-t)^4$, the result obtained from a product of dipole form factors consistent with the proton electric form factor.

This remains an interesting problem which deserves further study.


\subsection{The edge and diffraction dissociation \label{subset:diff_dissoc}}

An interesting connection between the edge and diffraction dissociation follows from an old analysis of the latter by Pumplin \cite{pumplin}, who used an argument based on unitarity and the properties of scattering eigenstates to show that the $b$-dependent partial cross sections for  the dissociation of an incident particle  on a nucleus  were bounded above by $(1/2)\sigma_{\rm tot}(b) - \sigma_{\rm elas}(b)$ where the cross sections refer to particle-nucleus scattering.  In the present case of strongly absorptive $pp$ or $\bar{p}p$ scattering, this argument leads to an upper bound on the single-particle dissociation cross section for either incident particle,
\ba
\label{sigma_SD}
\sigma_{\rm SD} &\leq& (\sigma_{\rm tot}-2\sigma_{\rm elas})/2 \\
\label{diss_xsec}
&=& 2\pi\int_0^\infty db\,b\eta(1-\eta),
\ea
or a total  dissociation cross section $\sigma_{\rm diss}\leq \sigma_{\rm tot}-2\sigma_{\rm elas}$ when both possibilities are included. The partial cross sections --- the sum of squares of the dissociation amplitudes --- are similarly bounded by the integrand in \eq{diss_xsec},
$\sigma_{SD}(b)\leq \eta(1-\eta)$.

The expression in \eq{diss_xsec} is just the edge integral, so the Pumplin bound relates $\sigma_{SD}$ to the area associated with the rim of width $t_{\rm edge}$ in the $pp$ or $\bar{p}p$ scattering amplitude.  Since  $t_{\rm edge}$  is essentially constant as seen in \fig{fig:edgewidthRcomp} and the edge integrand is  centered on $R_{\rm tot}=\sqrt{\sigma_{\rm tot}/2\pi}$, the bound on $\sigma_{\rm SD}$ grows proportionally to $\sqrt{\sigma_{\rm tot}}$ and will increase logarithmically at large $s$.
\be
\label{diss_xsec_growth}
\sigma_{\rm SD} \underset{s\rightarrow\infty}{\leq}{\rm constant}\times \ln{s},
\ee
a result which follows from the established $\ln^2s$ growth of $\sigma_{\rm tot}$ for $s\rightarrow\infty$.

Because \eq{diss_xsec_growth} only gives an upper bound on the growth of $\sigma_{SD}$ with energy, and there is no comparable energy-dependent lower bound, it is not clear that the real dissociation cross section will actually grow at this rate; however, that seems likely given the growth of other hadronic cross sections. The possibility of this behavior was noted in \cite{gotsman} for a specific model of diffraction dissociation, but some earlier models predicted a decreasing dissociation cross section.

Given the observed constancy of the edge width and the bound in \eq{sigma_SD}, we  find that   the ratio of the dissociation cross section to the total cross section must decrease at least logarithmically at high energies, $\sigma_{\rm SD}/\sigma_{\rm tot}\leq{\rm constant}\times 1/R_{\rm tot} \propto 1/\ln{s}\rightarrow 0$ for $s\rightarrow\infty$.

 As emphasized in \cite{gotsman}, many studies of particular mechanisms for diffractive dissociation neglect aborptive effects. These are clearly crucial in \eq{diss_xsec}: $\eta$ is very small at high energies out to impact parameters $b$ near $R_{\rm tot}$. Any reasonable model of diffractive dissociation must take this into account.

\begin{figure}[htbp]
\includegraphics{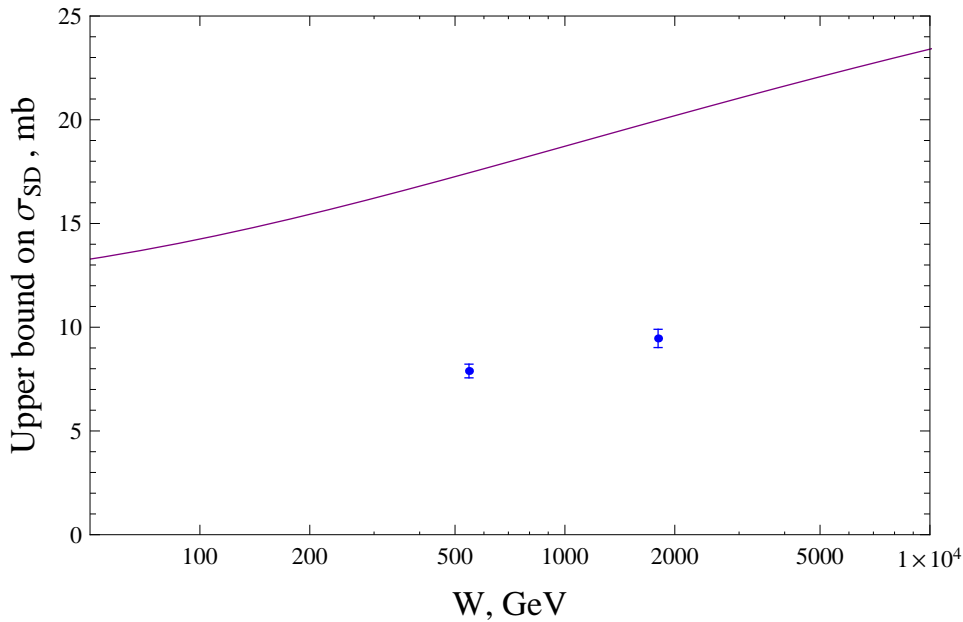}
 \caption{The curve gives the upper bound on the inclusive cross section for single-particle diffractive dissociation $\bar{p}+p\rightarrow \bar{p}+X$ calculated using $\sigma_{\rm SD}\leq(\sigma_{\rm tot}-2\sigma_{\rm elas})/2$ and our eikonal fit  to the $pp$ and $\bar{p}p$ cross section data from 5 GeV to 70 TeV. The data shown for $\sigma_{SD}$ are from CDF \cite{abe}. }
   \label{sigmadiffpbarp}
\end{figure}

We have calculated the bound on the single particle dissociation cross section using the expression in \eq{diss_xsec} and the eikonal factor $\eta(b,s)$ found in our fit to the combined $pp$ and $\bar{p}p$ data. We show the result and the CDF \cite{abe} measurements of diffractive dissociation in this process in \fig{sigmadiffpbarp}.

%

As expected, the measured cross section for diffractive dissociation lies considerably below the calculated inclusive value: the upper limit on $\sigma_{SD}$ is only reached under special conditions \cite{pumplin}. Furthermore, most experiments concentrate on the  differential cross section $d^2\sigma_{\rm SD}/dt\,dM_x^2$ in order to test specific models. The kinematic regions in which this can be measured and distinguished experimentally from other inelastic processes, are quite limited.  It is typical to require, for example, a very large ratio of the final center-of-mass momentum $p'$ of the surviving particle to its initial momentum $p$, $p'/p\gtrsim 0.85$ in the CDF experiments \cite{abe}, and further conditions on the detectability and mass of the dissociated system. There appear, in fact, to be no universally accepted experimental criteria for extracting this cross section, with the results typically depending on how the distributions in the momentum transfer $t$ and $M_X^2$ are modeled. The CDF results integrated over the allowed regions give cross sections which do increase  with energy roughly as predicted by the bound as shown in \fig{sigmadiffpbarp}.

Finally, to get an idea of the expected dependence of the scattering on the momentum transfer, we have calculated the analog of the elastic scattering amplitude $f(s,t)$, \eq{f}, for the exclusive process $\bar{p}+p\rightarrow \bar{p}+X$ using the bound $\eta(1-\eta)$ on the $b$-dependent amplitude noted above,
\be
\label{f_diff}
f_{\rm SD}(s,q) = \int_0^\infty db\,b\,\eta(1-\eta) J_0(qb).
\ee
Here $q^2=2pp'(1-\cos{\theta})=-t-(M_X^2-m^2)/2+\cdots$, $W\gg M_X^2, m^2$, where $\theta$ is angle through which the  the surviving particle is scattered and $M_X$ is the mass of the system $X$ \footnote{The Bessel function $J_0(qb)$ results from the conversion of the partial-wave series for the scattering amplitude to an integral over impact parameters using the approximate relation $P_j(\cos{\theta})\approx J_0(j\sqrt{2(1-\cos{\theta})})$, valid for small angles and large angular momenta $j$, the relation $q^2=2pp'(1-\cos{\theta})$, and the definition $b=j/\sqrt{pp'}$ for the impact parameter. This is the geometrical mean of the initial and final impact parameters.}. We do not specify the dependence of the dissociation process on $M_X$, but think of $f_{\rm SD}$ as giving the characteristic $b$ dependence of the mass-dependent amplitudes averaged over masses.  The
slope $B_{\rm SD}$ of the corresponding cross section at $\theta=0$ is defined for purely absorptive scattering ($\chi_R=0$) as
\be
\label{B_diss}
B_{\rm SD} = d\ln{|f_{\rm SD}|}^2/d\,q^2 = (1/2)\int_0^\infty db\,b^3\eta(1-\eta)\Big/\int_0^\infty db\,b\eta(1-\eta).
\ee

\begin{figure}[htbp]
\includegraphics{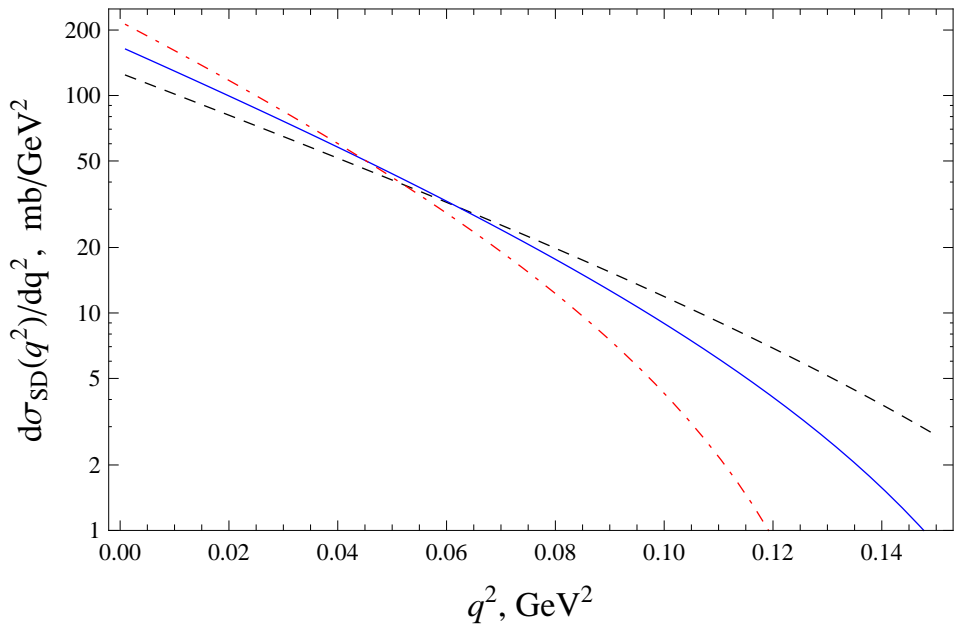}
\caption{The model differential dissociation cross sections calculated as the squares of the amplitudes $f_{\rm SD}$ in \eq{f_diff} for $W=546$ GeV (black dashed curve), 1800 GeV (solid blue curve), and 7000 GeV (dot-dashed red curve). The results illustrate the $q^2$  and $W$ dependence expected for dissociation ampitudes which saturate the edge distribution in impact parameter space, but are not predictions for the actual $M_X$-dependent cross sections.}
\label{fig:sigmadissq2}
\end{figure}

The results we obtain for $\left|f_{\rm SD}\right|^2$ are shown at several energies are in \fig{fig:sigmadissq2}.

%

We find that the cross section corresponding to the smooth peripheral edge distribution in \eq{diss_xsec} develops diffraction zeros at a given energy $W$ at smaller values of $q^2$ than the elastic cross section and that the forward slope parameter is larger, $B_{\rm SD}>B_{\rm elas}$. For example, at $W=1000$ GeV, $B_{\rm SD}=23.9$ GeV$^{-2}$ while $B_{\rm elas}=16.0$ GeV$^{-2}$. At 100 GeV, $B_{\rm SD}=18.0$ GeV$^{-2}$ while $B_{\rm elas}=13.3$ GeV$^{-2}$.  We emphasize that these results assume that the specific mechanism in question for  $\bar{p}+p\rightarrow \bar{p}+X$ is  represented in impact parameter space by a distribution $\propto\eta\left(1-\eta\right)$ which covers most of the region allowed by \eq{diss_xsec} at the given energy.

In the opposite  extreme in which the  mechanism for the dissociative production of a particular system $X$ is represented by a narrow distribution around an impact parameter $b_0$ in the allowed region,  $f_{\rm SD}\propto J_0(qb_0)$, and $B_{\rm SD}=b_0^2/2$. For $b_0$ near the peak of the edge distribution, $b_0\approx R_{\rm tot}=\sqrt{\sigma_{\rm tot}/2\pi}$, the slope parameter is somewhat smaller than $B_{\rm elas}$, 13.9 GeV$^{-2}$ instead of 16 GeV$^{-2}$ at $W=1000$ GeV, and 9.3 GeV$^{-2}$  versus 13.3 GeV$^{-2}$ at $W=100$ GeV. This picture is general because of the compact nature of the dissociation distribution in \eq{diss_xsec}. Any model for, or measurement of, diffraction dissociation must give a slope parameter in the range spanned by these limiting cases, most likely slightly larger than $R_{\rm tot}^2/2$ because  the extra factor of $b^2$ in the numerator of \eq{B_diss} weights that distribution toward larger impact parameters than that in \eq{diss_xsec} which is centered at $\approx R_{\rm tot}$.

Finally, we emphasize that it would be of considerable interest to measure the total single-dissociation cross section $\sigma_{\rm SD}$, for example for $\bar{p}+p\rightarrow \bar{p}+X$, without strong restrictions on $p'$ and $M_X$ and only the requirement that the final state contain an isolated $\bar{p}$ near the forward direction at a fairly low momentum transfer $q$, opposite a multi particle system. This would determine how close the bound on $\sigma_{SD}$ is to saturation.


\subsection{Survival of rapidity gaps \label{subsec:rapidity_gaps}}

The search for new physics in $pp$ or $\bar{p}p$ collisions can be simplified when the new process occurs in a large rapidity gap so is not accompanied by unrelated secondary particles in that region. An example discussed by Block and Halzen \cite{BHrapidity} would be Higgs boson production through $W$ boson fusion, $WW\rightarrow H$, where the $W$\,s are emitted by quarks in the colliding hadrons.

We take the inclusive differential cross section for this process in impact parameter space as
\be
\label{qqtoH}
\frac{d\sigma}{d^2{\bf b}} = \sigma_{WW\rightarrow H} A(b,\mu_{qq})
\ee
where $A(b,\mu_{qq})$ describes the spatial overlap of the quarks distributions as defined in Eqs.\ (\ref{overlap}) and (\ref{A(b,lambda)}).

$A(b,\mu_{qq})$ is normalized so that integration over $d^2{\bf b}$ with no further input would just give $\sigma_{WW\rightarrow H}$ as calculated in the parton model. However, further inelastic processes can occur in the hadronic collision giving secondary particles other than those associated with the remnants of the incident particles, and eliminating the rapidity gap. The probability that no such inelastic process occurs is $e^{-2\chi_I}$, and the cross section including this survival probability is therefore
\be
\label{qqtoHmod}
\frac{d\sigma}{d^2{\bf b}} = \sigma_{WW\rightarrow H} A(b,\mu_{qq}) e^{-2\chi_I(b,s)}.
\ee
The factor $A(b,\mu_{qq})e^{-2\chi_I(b,s)}$ is just the differential survival partiality for the gap. The construction generalizes to other processes.

Defining this following \cite{BHrapidity} as $d(\left|S\right|^2)/d^2{\bf b}$, the total survival probability for the gap is
\be
\langle\left|S\right|^2\rangle =  \int d^2{\bf b}\,A(b,\mu_{qq})e^{-2\chi_i(b,s)}.
\ee
These survival probabilities were calculated in \cite{BHrapidity} for the eikonal model discussed there. We have recalculated the survival probabilities using the eikonal model developed here. The results, given in Table \ref{tab:survival}, are very similar. The same calculation is easily done for gluon-initiated processes.


\begin{table}
\renewcommand{\arraystretch}{1.2}
 \caption{The gap survival probabilities $\langle\left|S\right|^2\rangle$ in percent for $pp$ and $\bar{p}p$ collisions as functions of the center-of-mass energy $W$.}
 \label{tab:survival}
\begin{tabular}{| l c  c c c  |}
\hline
\hline
$W$, GeV & & $pp\ (\%)$ & & $\bar{p}p\ (\%)$ \\
\hline
63 & & $38.7\pm 0.6\ \ $ & & $38.4\pm 0.6$ \\
546 & & $28.6\pm 0.5$ & & $28.6\pm 0.5$ \\
630 & & $27.8\pm 0.5 $ & & $27.8\pm 0.5$ \\
1,800 & & $22.2\pm 0.5$ & & $22.2\pm 0.5$ \\
14.000 & & $ 13.1\pm 0.3$ & & $ 13.1\pm 0.3$ \\
40,000 & & $9.8\pm 0.2$ & & $9.8\pm 0.2$\\
  \hline
\hline
 \end{tabular}
 \renewcommand{\arraystretch}{1}
 \end{table}



\section{Summary \label{sec:summary}}

In this paper, we presented the results of a detailed analysis of the current data on $pp$ and $\bar{p}p$ scattering in the eikonal formalism, parametrizing the eikonal function in a form suggested, but not restricted, by the structure found in minijet models for the scattering. The fit to the combined data is excellent. Our results for the total and elastic cross sections and $\rho$ values agree very well with the earlier fits of Block and Halzen \cite{blockhalzenfit,blockhalzen,blockrev,blockhalzen2} based on analytic amplitudes with a $\ln^2s$ growth at high energies.

We showed that our model, which includes a gluonic contribution  to the eikonal  function that grows as a power of $s=W^2$, leads naturally to cross sections, $\rho$ values, and slope parameters $B$ which can be described very accurately at high energies by quadratic expressions in $\ln{s}$. We can therefore extend the earlier Block-Halzen analysis of the energy dependence of the total cross sections and $\rho$ values to include the elastic and inelastic scattering cross sections and the logarithmic slope $B$ of the forward elastic scattering cross section, all measured quantities.

Our detailed model allowed us to analyze  the impact-parameter structure of the various scattering amplitudes in detail, including the relative importance of various contributions to the amplitudes, and the approach  to asymptotic behavior at high energies where gluonic processes become dominant. We commented on uncertainties in the asymptotic behavior of the eikonal function which affect the asymptotic behavior of the cross sections and merit further study.

We used our model to examine the structure of the edge of the $pp$ and $\bar{p}p$ scattering amplitudes recently identified by Block {\em et al.}  \cite{edge} in some detail.  The width of this edge region is nearly energy independent at $\sim 1$ fm, a property clearly evident in our results. We commented on some possible dynamical origins for the edge.

We also used the model to investigate the Pumplin bound \cite{pumplin} on the cross section for single particle diffractive dissociation which is given directly in terms of the edge cross section. The constancy of $t_{\rm edge}$ and the $\ln^2s$ growth of $\sigma_{\rm tot}$ at high energies show that the bound --- and possibly $\sigma_{SD}$ --- increase only as $\ln{s}$, while the ratio $\sigma_{SD}/\sigma_{\rm tot}$ must decrease at least as $1/\ln{s}$.

Finally, we used the model to update earlier results \cite{BHrapidity} on the survival probability of large rapidity gaps in $pp$ and $\bar{p}p$ scattering, a matter of interest in the search for rare processes in the scattering.


\begin{acknowledgments}

M.M.B., L.D., and F.H.\  would  like to thank the Aspen Center for Physics for its hospitality and for its partial support of this work under NSF Grant No. 1066293. F.H.'s research was supported in part by the U.S. National Science
Foundation under Grants No.~OPP-0236449 and PHY-0969061 and by the
University of Wisconsin Research Committee with funds granted by the Wisconsin Alumni Research
Foundation.   P.H.\ would like to thank Towson University Fisher College of Science and Mathematics for support.   The authors would like to thank Leo Stodolsky and Thomas J.\ Weiler for extensive discussions about the edge in the original stages of this work.

\end{acknowledgments}


\appendix

\section{The modified Aspen model \label{model} }

The model used here in fitting the $pp$ and $p\bar{p}$ cross sections is a modification of the ``Aspen model'' of  Block {\em et al.\/} \cite{blockaspen,blockrev}. That model was based the structure of the eikonal function found in QCD minijet models for the scattering in which the interactions between hadrons are described in terms of the interactions of their constituent quarks and gluons with allowance for ``soft'' interactions at low momentum transfers. While we will follow the notation used in \cite{blockaspen},  the identification of the terms made there as describing quark-quark ($qq$), quark-gluon ($qg$), or gluon-gluon ($gg$) interactions becomes blurred in the general setting, especially for the $qq$ and $qg$ terms.

We will write the eikonal functions in terms of crossing-even and crossing-odd components as
\ba
\label{chiEmodel2}
\chi_E(b,W) &=& i\left[\sigma_{qq}(w) A(b,\mu_{qq}) + \sigma_{qg}(w) A(b,\mu_{qg}) + \sigma_{gg}(w) A(b,\mu_{gg})\right], \\
\label{chiOmodel2}
\chi_O(b,W) &=& -\Sigma_{gg}C_5\left(\frac{m_0}{w}\right)^{2-2\alpha_1} A(b,\mu_{odd}),
\ea
where one needs to make the replacement $w\rightarrow We^{-i\pi/4}$ in the final results to obtain the correct asymptotic phase required by  analyticity and crossing symmetry \cite{blockcahn,blockrev}. Here we will simply write the functions on the right-hand sides of \eq{chiEmodel2} and \eq{chiOmodel2} as functions of $w$, with the replacement to be made in the final results. The constant $\Sigma_{gg}=9\pi\alpha_s^2/m_0^2$ sets the scale in \eq{chiOmodel2} and later equations.

The overlap factors $A(b,\lambda)$ in these expressions  are defined in terms of the relevant distributions in the proton by
\be
\label{overlap}
A_{ij} = \int d^2{\mathbf b}\, \rho_i({\mathbf b'}) \rho_j(\mathbf{ b-b'}).
\ee
Assuming that the distributions $\rho_i$ have approximately the same form as that determined from the proton electric form factor, the overlap functions become
\be
\label{A(b,lambda)}
A(b,\lambda) = \frac{\lambda^2}{96\pi}(\lambda b)^3 K_3(\lambda b), \quad \int_0^\infty d^2{\mathbf b}A(b,\lambda)=1
\ee
for appropriate choices of the $\lambda$ parameters.

The gluon-gluon term in \eq{chiEmodel2}, dominant at very high energies, was parametrized in \cite{blockaspen}  using a very simplified description of $gg$ scattering in low-order QCD.  The result was an expression which involved a leading power of $s/m_0^2$, logarithms of that quantity, and a constant term, plus terms involving inverse powers of $s/m_0^2$. Given the uncertainties in the model, including a rather arbitrary choice of the leading power, we will simply parametrize $\sigma_{gg}$ directly in terms of a power and leading logarithm in $s/m_0^2$ with an additive constant chosen so that the $gg$ term gives a negligible contribution to the eikonal function at low energies as in low-order QCD.

The $qg$ and $qq$ terms have a less singular structure in QCD, and model results derived using scaling parametrization of the quark structure function $f_q$ do not separate cleanly from the expected contributions from soft processes or the $gg$ terms. We simply follow the parametrizations used in \cite{blockaspen} allowing, however, the powers in the Regge-like low-energy terms to vary from the $1/\sqrt{s}=1/W$ behavior assumed there.

The ``cross sections'' $\sigma_{ij}$ in \eq{chiEmodel2} are then
\ba
\label{sigma_qq}
\sigma_{qq}(w) &=&  \Sigma_{gg}\left[C_0 + C_1(m_0/w)^{2 - 2 \alpha_2}\right], \\
\label{sigma_qg}
\sigma_{qg}(w) &=&  \Sigma_{gg}C_2\ln(w^2/m_0^2),  \\
\sigma_{gg}(w) &=&  \Sigma_{gg}\left\{0.0713+C_3\ln(W/W_0)+C_4\left[(W/m_0)^\beta-(W_0/m_0)^\beta\right]\right\}
\label{sigma_gg}
\ea
The expression in \eq{sigma_gg} gives an excellent fit to the more complicated and restricted form for $\sigma_{gg}$ derived in \cite{blockaspen} at higher energies where this term is important,  and contributes less than 1\% of the total eikonal function at $b=0$ at the low-energy matching point $W=W_0=4$ GeV in agreement with the results there.

Our fit to the $pp$ and $\bar{p}p$ scattering cross sections, the ratios $\rho$ of the real to the imaginary part of the forward elastic scattering amplitudes, and the logarithmic derivatives $B$  of the forward differential cross sections $d\sigma/dt$ used the six coefficients $C_0,\,C_1,\,C_2,\,C_3,\,C_4,\,C_5$ and the parameters $\alpha_1$, $\alpha_2$, and $\beta$.
The remaining parameters $\mu_{gg},\,\mu_{qq}$ and $m_0$ were fixed as in \cite{blockaspen} with the energy scale $m_0=0.6$ GeV, and the $\mu$\,s chosen by hand in the range determined by the proton charge form factor, $\mu_{gg}=0.705$ GeV and $\mu_{qq}=0.89$ GeV. We did not vary these parameters in making the fit. We note also that the overall factor  $\Sigma_{gg}=9\pi\alpha_s^2/m_0^2$ which appears in the cross sections in \eq{chiOmodel2} and Eqs.\ (\ref{sigma_qq}-\ref{sigma_qq}) can be absorbed into the coefficients $C_i$; it was separated out in \cite{blockhalzen,blockrev}  to provide a connection with minijet models for the eikonal function $\chi$ where such factors appear naturally.

A summary of the parameters with the results of the fit is  given in Table \ref{tab:parameters}.


\begin{table}
\renewcommand{\arraystretch}{1.2}
 \caption{Summary of the parameters used in the fit to the $pp$ and $\bar{p}p$ scattering data}
 \label{tab:parameters}
\begin{tabular}{| c c l |}
\hline
\hline
Fixed values && Fitted parameters \\
\hline
$m_0=0.6$ GeV && $C_0=6.086 \pm 0.07$ \\
$W_0=4$ GeV && $C_1=29.22 \pm 0.02$\\
$\mu_{gg}=0.705$ GeV && $C_2=0.0130 \pm 0.0004 $\\
$\mu_{qq}= 0.89 $ GeV && $C_3=-2.258 \pm 0.004 $ \\
$\mu_{odd}= 0.60$ GeV && $C_4=8.762 \pm 0.013 $ \\
&& $C_5=-26.206 \pm 0.02$ \\
$\alpha_s=0.5$ && $\alpha_1=0.3171 \pm 0.0003 $ \\
 $\Sigma_{gg}=9\pi\alpha_s^2/m_0^2$ && $\alpha_2= 0.4606 \pm 0.0001 $ \\
$\  =19.635$ GeV$^{-2}$  && $\beta=0.1726\pm 0.0002$ \\
  \hline
\hline
 \end{tabular}
 \renewcommand{\arraystretch}{1}
 \end{table}


\bibliography{small_x_references}

\begin{thebibliography}{31}
\expandafter\ifx\csname natexlab\endcsname\relax\def\natexlab#1{#1}\fi
\expandafter\ifx\csname bibnamefont\endcsname\relax
  \def\bibnamefont#1{#1}\fi
\expandafter\ifx\csname bibfnamefont\endcsname\relax
  \def\bibfnamefont#1{#1}\fi
\expandafter\ifx\csname citenamefont\endcsname\relax
  \def\citenamefont#1{#1}\fi
\expandafter\ifx\csname url\endcsname\relax
  \def\url#1{\texttt{#1}}\fi
\expandafter\ifx\csname urlprefix\endcsname\relax\def\urlprefix{URL }\fi
\providecommand{\bibinfo}[2]{#2}
\providecommand{\eprint}[2][]{\url{#2}}

\bibitem[{\citenamefont{Block et~al.}(2015)\citenamefont{Block, Durand, Halzen,
  Stodolsky, and Weiler}}]{edge}
\bibinfo{author}{\bibfnamefont{M.~M.} \bibnamefont{Block}},
  \bibinfo{author}{\bibfnamefont{L.}~\bibnamefont{Durand}},
  \bibinfo{author}{\bibfnamefont{F.}~\bibnamefont{Halzen}},
  \bibinfo{author}{\bibfnamefont{L.}~\bibnamefont{Stodolsky}},
  \bibnamefont{and} \bibinfo{author}{\bibfnamefont{T.}~\bibnamefont{Weiler}},
  \bibinfo{journal}{Phys. Rev. D} \textbf{\bibinfo{volume}{91}},
  \bibinfo{pages}{011501(R)} (\bibinfo{year}{2015}), \eprint{arXiv:1409.3196
  [hep-ph]}.

\bibitem[{\citenamefont{Block and Halzen}(2005)}]{blockhalzenfit}
\bibinfo{author}{\bibfnamefont{M.~M.} \bibnamefont{Block}} \bibnamefont{and}
  \bibinfo{author}{\bibfnamefont{F.}~\bibnamefont{Halzen}},
  \bibinfo{journal}{Phys. Rev. D} \textbf{\bibinfo{volume}{72}},
  \bibinfo{pages}{036006} (\bibinfo{year}{2005}).

\bibitem[{\citenamefont{Block and Halzen}(2011)}]{blockhalzen}
\bibinfo{author}{\bibfnamefont{M.~M.} \bibnamefont{Block}} \bibnamefont{and}
  \bibinfo{author}{\bibfnamefont{F.}~\bibnamefont{Halzen}},
  \bibinfo{journal}{Phys. Rev. Lett.} \textbf{\bibinfo{volume}{107}},
  \bibinfo{pages}{212002} (\bibinfo{year}{2011}).

\bibitem[{\citenamefont{Block and Halzen}(2012)}]{blockhalzen2}
\bibinfo{author}{\bibfnamefont{M.~M.} \bibnamefont{Block}} \bibnamefont{and}
  \bibinfo{author}{\bibfnamefont{F.}~\bibnamefont{Halzen}},
  \bibinfo{journal}{Phys. Rev. D} \textbf{\bibinfo{volume}{86}},
  \bibinfo{pages}{051504} (\bibinfo{year}{2012}).

\bibitem[{\citenamefont{Fagundes et~al.}(2015)\citenamefont{Fagundes, Grau,
  Pancheri, Srivastava, and Shekhovtsova}}]{fagundes}
\bibinfo{author}{\bibfnamefont{D.~A.} \bibnamefont{Fagundes}},
  \bibinfo{author}{\bibfnamefont{A.}~\bibnamefont{Grau}},
  \bibinfo{author}{\bibfnamefont{G.}~\bibnamefont{Pancheri}},
  \bibinfo{author}{\bibfnamefont{Y.~N.} \bibnamefont{Srivastava}},
  \bibnamefont{and}
  \bibinfo{author}{\bibfnamefont{O.}~\bibnamefont{Shekhovtsova}},
  \bibinfo{journal}{Phys. Rev. D} \textbf{\bibinfo{volume}{91}},
  \bibinfo{pages}{114011} (\bibinfo{year}{2015}).

\bibitem[{\citenamefont{Khoze et~al.}(2015)\citenamefont{Khoze, Martin, and
  Ryskin}}]{KhozeMartinRyskin}
\bibinfo{author}{\bibfnamefont{V.~A.} \bibnamefont{Khoze}},
  \bibinfo{author}{\bibfnamefont{A.~D.} \bibnamefont{Martin}},
  \bibnamefont{and} \bibinfo{author}{\bibfnamefont{M.~G.}
  \bibnamefont{Ryskin}}, \bibinfo{journal}{Intl. J. Mod. Phys. A}
  \textbf{\bibinfo{volume}{30}}, \bibinfo{pages}{1542004}
  (\bibinfo{year}{2015}).

\bibitem[{\citenamefont{Gotsman et~al.}(2015)\citenamefont{Gotsman, Levin, and
  Maor}}]{GotsmanLHC}
\bibinfo{author}{\bibfnamefont{E.}~\bibnamefont{Gotsman}},
  \bibinfo{author}{\bibfnamefont{E.}~\bibnamefont{Levin}}, \bibnamefont{and}
  \bibinfo{author}{\bibfnamefont{U.}~\bibnamefont{Maor}},
  \bibinfo{journal}{Intl. J. Mod. Phys. A} \textbf{\bibinfo{volume}{30}},
  \bibinfo{pages}{154205} (\bibinfo{year}{2015}).

\bibitem[{\citenamefont{Pumplin}(1973)}]{pumplin}
\bibinfo{author}{\bibfnamefont{J.}~\bibnamefont{Pumplin}},
  \bibinfo{journal}{Phys. Rev. D} \textbf{\bibinfo{volume}{8}},
  \bibinfo{pages}{2899} (\bibinfo{year}{1973}).

\bibitem[{\citenamefont{Block}(2006{\natexlab{a}})}]{blockrev}
\bibinfo{author}{\bibfnamefont{M.~M.} \bibnamefont{Block}},
  \bibinfo{journal}{Phys. Rep.} \textbf{\bibinfo{volume}{436}},
  \bibinfo{pages}{71} (\bibinfo{year}{2006}{\natexlab{a}}).

\bibitem[{\citenamefont{Block et~al.}(1999)\citenamefont{Block, Gregores,
  Halzen, and Pancheri}}]{blockaspen}
\bibinfo{author}{\bibfnamefont{M.~M.} \bibnamefont{Block}},
  \bibinfo{author}{\bibfnamefont{E.~M.} \bibnamefont{Gregores}},
  \bibinfo{author}{\bibfnamefont{F.}~\bibnamefont{Halzen}}, \bibnamefont{and}
  \bibinfo{author}{\bibfnamefont{G.}~\bibnamefont{Pancheri}},
  \bibinfo{journal}{Phys. Rev. D} \textbf{\bibinfo{volume}{60}},
  \bibinfo{pages}{054024} (\bibinfo{year}{1999}).

\bibitem[{\citenamefont{Block and Cahn}(1985)}]{blockcahn}
\bibinfo{author}{\bibfnamefont{M.~M.} \bibnamefont{Block}} \bibnamefont{and}
  \bibinfo{author}{\bibfnamefont{R.~N.} \bibnamefont{Cahn}},
  \bibinfo{journal}{Rev. Mod. Phys.} \textbf{\bibinfo{volume}{57}},
  \bibinfo{pages}{563} (\bibinfo{year}{1985}).

\bibitem[{\citenamefont{{ATLAS Collaboration}}(2011)}]{LHCtot1}
\bibinfo{author}{\bibnamefont{{ATLAS Collaboration}}}, \bibinfo{journal}{Nature
  Comm.} \textbf{\bibinfo{volume}{2}}, \bibinfo{pages}{463}
  (\bibinfo{year}{2011}).

\bibitem[{\citenamefont{Antchev et~al.}(2011{\natexlab{a}})}]{LHCtot2}
\bibinfo{author}{\bibfnamefont{G.}~\bibnamefont{Antchev}} \bibnamefont{et~al.}
  (\bibinfo{collaboration}{TOTEM Collaboration}), \bibinfo{journal}{Euro. Phys.
  Lett.} \textbf{\bibinfo{volume}{96}}, \bibinfo{pages}{21002}
  (\bibinfo{year}{2011}{\natexlab{a}}).

\bibitem[{LHC()}]{LHCtot3}
\bibinfo{note}{CMS Collaboration, \uppercase{C}ERN Document Server,
  http://cdsweb.cern.ch/record/1373466?ln=en, 2011}.

\bibitem[{\citenamefont{Antchev et~al.}(2013{\natexlab{a}})}]{totem2013}
\bibinfo{author}{\bibfnamefont{G.}~\bibnamefont{Antchev}} \bibnamefont{et~al.}
  (\bibinfo{collaboration}{TOTEM Collaboration}), \bibinfo{journal}{Euro. Phys.
  Lett.} \textbf{\bibinfo{volume}{101}}, \bibinfo{pages}{21002}
  (\bibinfo{year}{2013}{\natexlab{a}}).

\bibitem[{\citenamefont{Antchev et~al.}(2013{\natexlab{b}})}]{totem2013_2}
\bibinfo{author}{\bibfnamefont{G.}~\bibnamefont{Antchev}} \bibnamefont{et~al.}
  (\bibinfo{collaboration}{TOTEM Collaboration}), \bibinfo{journal}{Euro.Phys.
  Lett.} \textbf{\bibinfo{volume}{101}}, \bibinfo{pages}{21004}
  (\bibinfo{year}{2013}{\natexlab{b}}).

\bibitem[{\citenamefont{Abreu et~al.}(2012)}]{POAp-air}
\bibinfo{author}{\bibfnamefont{P.}~\bibnamefont{Abreu}} \bibnamefont{et~al.}
  (\bibinfo{collaboration}{Pierre Auger Collaboration}),
  \bibinfo{journal}{Phys. Rev. Lett.} \textbf{\bibinfo{volume}{109}},
  \bibinfo{pages}{062002} (\bibinfo{year}{2012}), \eprint{arXiv:1208.1520
  [hep-ex]}.

\bibitem[{\citenamefont{Abbasi et~al.}(2008)}]{HiRes}
\bibinfo{author}{\bibfnamefont{R.}~\bibnamefont{Abbasi}} \bibnamefont{et~al.}
  (\bibinfo{collaboration}{HiRes Collaboration}), \bibinfo{journal}{Ap. J.}
  \textbf{\bibinfo{volume}{684}}, \bibinfo{pages}{790} (\bibinfo{year}{2008}).

\bibitem[{\citenamefont{Block}(2006{\natexlab{b}})}]{sieve}
\bibinfo{author}{\bibfnamefont{M.~M.} \bibnamefont{Block}},
  \bibinfo{journal}{Nucl. Inst. and Meth. A.} \textbf{\bibinfo{volume}{556}},
  \bibinfo{pages}{308} (\bibinfo{year}{2006}{\natexlab{b}}).

\bibitem[{\citenamefont{Amos et~al.}(1988)}]{E710_B}
\bibinfo{author}{\bibfnamefont{N.}~\bibnamefont{Amos}} \bibnamefont{et~al.}
  (\bibinfo{collaboration}{E710 Collaboration}), \bibinfo{journal}{Phys. Rev.
  Lett.} \textbf{\bibinfo{volume}{61}}, \bibinfo{pages}{525}
  (\bibinfo{year}{1988}).

\bibitem[{\citenamefont{Amos et~al.}(1989)}]{E710_B2}
\bibinfo{author}{\bibfnamefont{N.}~\bibnamefont{Amos}} \bibnamefont{et~al.}
  (\bibinfo{collaboration}{E710 Collaboration}), \bibinfo{journal}{Phys. Rev.
  Lett.} \textbf{\bibinfo{volume}{63}}, \bibinfo{pages}{2784}
  (\bibinfo{year}{1989}).

\bibitem[{\citenamefont{Antchev et~al.}(2011{\natexlab{b}})}]{totem2011}
\bibinfo{author}{\bibfnamefont{G.}~\bibnamefont{Antchev}} \bibnamefont{et~al.}
  (\bibinfo{collaboration}{TOTEM Collaboration}), \bibinfo{journal}{Euro. Phys.
  Lett.} \textbf{\bibinfo{volume}{95}}, \bibinfo{pages}{41001}
  (\bibinfo{year}{2011}{\natexlab{b}}).

\bibitem[{\citenamefont{Schegelsky and Ryskin}(2012)}]{ryskin}
\bibinfo{author}{\bibfnamefont{V.~A.} \bibnamefont{Schegelsky}}
  \bibnamefont{and} \bibinfo{author}{\bibfnamefont{M.~G.}
  \bibnamefont{Ryskin}}, \bibinfo{journal}{Phys. Rev. D}
  \textbf{\bibinfo{volume}{85}}, \bibinfo{pages}{094024}
  (\bibinfo{year}{2012}), \eprint{arXiv:1112.3243 [hep-ph]}.

\bibitem[{\citenamefont{Olive et~al.}(2104)}]{PDG2014}
\bibinfo{author}{\bibfnamefont{K.~A.} \bibnamefont{Olive}} \bibnamefont{et~al.}
  (\bibinfo{collaboration}{Particle Data Group}), \bibinfo{journal}{Chin. Phys.
  C} \textbf{\bibinfo{volume}{38}}, \bibinfo{pages}{090001}
  (\bibinfo{year}{2104}), \bibinfo{note}{\uppercase{S}ec. 50, Total Hadronic
  Cross Sections}.

\bibitem[{\citenamefont{Froissart}(1961)}]{froissart}
\bibinfo{author}{\bibfnamefont{M.}~\bibnamefont{Froissart}},
  \bibinfo{journal}{Phys. Rev.} \textbf{\bibinfo{volume}{123}},
  \bibinfo{pages}{1053} (\bibinfo{year}{1961}).

\bibitem[{\citenamefont{Martin}(1963)}]{martin1}
\bibinfo{author}{\bibfnamefont{A.}~\bibnamefont{Martin}},
  \bibinfo{journal}{Phys. Rev.} \textbf{\bibinfo{volume}{129}},
  \bibinfo{pages}{1432} (\bibinfo{year}{1963}).

\bibitem[{\citenamefont{Rosner}(2014)}]{rosner}
\bibinfo{author}{\bibfnamefont{J.}~\bibnamefont{Rosner}},
  \bibinfo{journal}{Phys. Rev. D} \textbf{\bibinfo{volume}{90}},
  \bibinfo{pages}{117902} (\bibinfo{year}{2014}), \eprint{arXiv:1409.5813
  [hep-ph]}.

\bibitem[{\citenamefont{Wu and Yang}(1965)}]{wuyang}
\bibinfo{author}{\bibfnamefont{T.~T.} \bibnamefont{Wu}} \bibnamefont{and}
  \bibinfo{author}{\bibfnamefont{C.~N.} \bibnamefont{Yang}},
  \bibinfo{journal}{Phys. Rev. B} \textbf{\bibinfo{volume}{137}},
  \bibinfo{pages}{B708} (\bibinfo{year}{1965}).

\bibitem[{\citenamefont{Gotsman et~al.}(1994)\citenamefont{Gotsman, Levin, and
  Maor}}]{gotsman}
\bibinfo{author}{\bibfnamefont{E.}~\bibnamefont{Gotsman}},
  \bibinfo{author}{\bibfnamefont{E.~M.} \bibnamefont{Levin}}, \bibnamefont{and}
  \bibinfo{author}{\bibfnamefont{U.}~\bibnamefont{Maor}},
  \bibinfo{journal}{Phys. Rev. D} \textbf{\bibinfo{volume}{49}},
  \bibinfo{pages}{R4321} (\bibinfo{year}{1994}).

\bibitem[{\citenamefont{Abe et~al.}(1994)}]{abe}
\bibinfo{author}{\bibfnamefont{F.}~\bibnamefont{Abe}} \bibnamefont{et~al.}
  (\bibinfo{collaboration}{CDF Collaboration}), \bibinfo{journal}{Phys. Rev. D}
  \textbf{\bibinfo{volume}{50}}, \bibinfo{pages}{5535} (\bibinfo{year}{1994}).

\bibitem[{\citenamefont{Block and Halzen}(2001)}]{BHrapidity}
\bibinfo{author}{\bibfnamefont{M.~M.} \bibnamefont{Block}} \bibnamefont{and}
  \bibinfo{author}{\bibfnamefont{F.}~\bibnamefont{Halzen}},
  \bibinfo{journal}{Phys. Rev. D} \textbf{\bibinfo{volume}{63}},
  \bibinfo{pages}{114004} (\bibinfo{year}{2001}).

\end{thebibliography}

\end{document}